\documentclass[peerreview, 11pt]{IEEEtran}
\usepackage{cite,epsfig,graphicx, amsmath, amssymb, latexsym,subfigure, theorem}

\begin{document}
\title{Complementary Set Matrices Satisfying \\ a Column Correlation Constraint}
\author{Di Wu\thanks{* This work has been supported in part by the NSF Grant ANI 0338805} and Predrag Spasojevi\'c \\
WINLAB, Rutgers University \\
671 Route 1 South, North Brunswick, NJ 08902\\
\{diwu,spasojev\}@winlab.rutgers.edu} \maketitle
\pagestyle{empty} \thispagestyle{empty}

\begin{abstract}
Motivated by the problem of reducing the peak to average power
ratio (PAPR) of transmitted signals, we consider a design of
complementary set matrices whose column sequences satisfy a
correlation constraint. The design algorithm recursively builds a
collection of $2^{t+1}$ mutually orthogonal (MO) complementary set
matrices starting from a companion pair of sequences. We relate
correlation properties of column sequences to that of the
companion pair and illustrate how to select an appropriate
companion pair to ensure that a given column correlation
constraint is satisfied. For $t=0$, companion pair properties
directly determine matrix column correlation properties. For $t
\geq 1$, reducing correlation merits of the companion pair may
lead to improved column correlation properties. However, further
decrease of the maximum out-off-phase aperiodic autocorrelation of
column sequences is not possible once the companion pair
correlation merit is less than a threshold determined by $t$. We
also reveal a design of the companion pair which leads to
complementary set matrices with Golay column sequences. Exhaustive
search for companion pairs satisfying a column correlation
constraint is infeasible for medium and long sequences. We instead
search for two shorter length sequences by minimizing a cost
function in terms of their autocorrelation and crosscorrelation
merits. Furthermore, an improved cost function which helps in
reducing the maximum out-off-phase column correlation is derived
based on the properties of the companion pair. By exploiting the
well-known Welch bound, sufficient conditions for the existence of
companion pairs which satisfy a set of column correlation
constraints are also given.
\end{abstract}

\begin{keywords} Complementary sets, Golay sequences,
peak to average power ratio (PAPR), Welch bound
\end{keywords}

\section{Introduction}
Complementary sequence sets have been introduced by
Golay~\cite{Golay:49, Golay:61}, as a pair of binary sequences
with the property that the sum of their aperiodic autocorrelation
functions (ACF) is zero everywhere except at zero shift. Tseng and
Liu~\cite{Tseng:72} generalized these ideas to sets of binary
sequences of size larger than two. Sivaswamy~\cite{Siva:78} and
Frank~\cite{Frank:80} investigated the multiphase (polyphase)
complementary sequence sets with constant amplitude sequence
elements. Gavish and Lemple considered ternary complementary pairs
over the alphabet $\{1, 0,-1\}$~\cite{Gavish:94}. The synthesis of
multilevel complementary sequences is described
in~\cite{Darnell:88}. These generalizations of a binary alphabet
lead to new construction methods for complementary sets having a
larger family of lengths and cardinalities. However, all these
studies focus either on the set complementarity or on the design
of orthogonal families of complementary sets. Correlation
properties of column sequences of the complementary set matrix
(i.e., the matrix whose row sequences form a complementary set)
have not been considered.

In~\cite{Tseng:00, Di:03, chen:01}, a technique for the
multicarrier direct-sequence code-division multiple access
(MC-DS-CDMA) system~\cite{Kondo:96, Sourour:96} that employs
complementary sets as spreading sequences has been investigated.
Each user assigns different sequences from a complementary set to
his subcarriers. By assigning mutually orthogonal (MO)
complementary sets to different users, both multiple access
interference and multipath interference can be significantly
suppressed. Similar to conventional multicarrier systems, one of
the major impediments to deploying such systems is high
peak-to-average power ratio (PAPR). We have stressed
in~\cite{Di:04} that correlation properties of column sequences of
complementary set matrices play an important role in the reduction
of PAPR. In this work, we search for ways of constructing
complementary set matrices whose column sequences satisfy a
correlation constraint.

For orthogonal frequency-division multiplexing (OFDM) signaling,
Tellambura~\cite{Tell:97} derived a general upper bound on the
signal peak envelope power (PEP) in terms of the aperiodic ACF of
the sequence whose elements are assigned across all signal
carriers. He has shown that sequences with small aperiodic
autocorrelation values can reduce the PAPR of the OFDM signal. By
generalizing earlier work of Boyd~\cite{Boyd:86},
Popovi\'c~\cite{popovic:91} has demonstrated that PAPR
corresponding to any binary Golay sequence (i.e., a sequence
having a Golay complementary pair) is at most two. This has
motivated Davis and Jedwab to explicitly determine a large class
of Golay sequences as a solution to the signal envelope
problem~\cite{Davis:99}. Here, we consider sequence sets which are
characterized by both their complementarity and a desired column
correlation constraint.

We describe a construction algorithm for the design of $2^{t+1}$
MO complementary set matrices of size $2^tm$ by $2^{t+p+1}$ in
Section III, where $t$ and $p$ can be any non-negative integer,
and $m$ is an even number. The construction process is based on a
set of sequence/matrix operations, starting from a two column
matrix formed by a companion sequence pair. These operations
preserve the alphabet (up to the sign) of the companion pair. In
Section IV, we illustrate how, by selecting an appropriate
companion pair we can ensure that column sequences of the
constructed complementary set matrix satisfy a correlation
constraint. For $t=0$, companion pair properties directly
determine matrix column correlation properties. For $t \geq 1$,
reducing correlation merits of the companion pair may lead to
improved column correlation properties. However, further decrease
of the maximum out-off-phase aperiodic autocorrelation of column
sequences is not possible once the correlation of the companion
pair is less than a threshold determined by $t$. We also present a
method for constructing the companion pair which leads to the
complementary set matrix with Golay column sequences.

In Section V, an algorithm for searching for companion pairs over
length $m$ sequences of a desired alphabet is described. However,
exhaustive search is infeasible for medium and long sequences. We
instead suggest finding companion pairs with a small, if not
minimum, column correlation constraint. In Section VI, by
exploiting properties of the companion pair, we convert the
problem into a search for two sequences of length $m/2$ with low
autocorrelation and crosscorrelation merits, a long standing
problem in literature (e.g. see \cite{Somaini:75, Sarwate:79,
Hai:96, Mow:97}). We further derive an improved cost function and
show how it leads to reduced achievable maximum out-off-phase
column correlation constraint. Sufficient conditions for the
existence of companion pairs which satisfy various column
correlation constraints are also derived. We conclude in Section
VII.

\section{Definitions and Preliminaries}

Throughout this paper, sequences are denoted by boldface lowercase
letters (e.g., $\mathbf{x}$), their elements by corresponding
lowercase letters with subscripts (${x}_0$), boldface uppercase
letters denote matrices ($\mathbf{X}$), and calligraphic letters
denote either sets of numbers or sets of sequences (${\cal X}$).

\subsection{Correlation functions}
Let $\mathbf{a}= (a_0, a_1, ..., a_{n-1})$ denote a sequence of
length $n$ with $a_i \in \cal C$, $ 0 \leq i \leq n-1$, where
$\cal C$ is the set of complex numbers. The aperiodic and periodic
ACFs of $\mathbf{a}$ are
\begin{eqnarray}
A_{\mathbf{a}}(l)&=&\sum_{i=0}^{n-1-l} a_ia^*_{i+l}, \,\,\,\, 0
\leq l \leq n-1  
\\
P_{\mathbf{a}} (l) &=&  \sum^{n-1}_{i=0} a_ia^*_{i \, \oplus_n \,
l},\,\,\,\,\; 0 \leq l \leq n-1 ,
\end{eqnarray}
where $a^*$ denotes the complex conjugate of $a$, and $\oplus_n$
denotes modulo-n addition. It follows that
\begin{eqnarray}
P_{\mathbf{a}}(l) = A_{\mathbf{a}} (l) + A_{\mathbf{a}}
(n-l),\,\,\,\,\; 0 \leq l \leq n-1. \label{AtoP}
\end{eqnarray}

Let $\mathbf{b}= (b_0, b_1, ..., b_{n-1})$, where $b_i \in \cal
C$, $ 0 \leq i \leq n-1$. The aperiodic and periodic
crosscorrelation functions of $\mathbf{a}$ and $\mathbf{b}$ are
defined, respectively, as,
\begin{eqnarray}
A_{\mathbf{a},\mathbf{b}} (l) &=& \left \{ \begin {array} {cc}
\sum\limits^{n-1-l}_{i=0} a_ib^*_{i+l},\,\,\,\, & 0 \leq l \leq n-1 \\
\sum\limits^{n-1+l}_{i=0} a_{i-l}b^*_{i},\,\,\,\, & 1-n \leq l < 0 \\
0,\,\,\,\, & |l| \geq n \\
\end{array} \right. 
\\
P_{\mathbf{a},\mathbf{b}} (l) &=&  \sum^{n-1}_{i=0} a_ib^*_{i \,
\oplus_n \, l},\,\,\,\,\; 0 \leq l \leq n-1 \;. 
\end{eqnarray}

Table~I lists the correlation function parameters which are
commonly used to judge the merits of a sequence design, and are
termed \textit{correlation merits}. $\lambda^{A}_{\mathbf{a}}$,
$\lambda^{P}_{\mathbf{a}}$, $S^{A}_{\mathbf{a}}$, and
$S^{P}_{\mathbf{a}}$ are autocorrelation merits, and
$\lambda^{A}_{\mathbf{a,b}}$, $\lambda^{P}_{\mathbf{a,b}}$,
$S^{A}_{\mathbf{a,b}}$, and $S^{P}_{\mathbf{a,b}}$ are common
crosscorrelation merits. For example, it is well known that binary
$m$-sequences satisfy $\lambda^{P}_{\mathbf{a}} = 1$, and the
sequences with $\lambda^{A}_{\mathbf{a}} \leq 1$ are called Barker
sequences. Furthermore, a small value of $S^{A}_{\mathbf{a}}$ can
significantly reduce the PAPR of OFDM signals, if the elements of
$\mathbf{a}$ are assigned across all
carriers~\cite{Di:04,Tell:97}.


\begin{table} [!t]
\begin{center}\caption{Correlation Merits} {
\begin{tabular}{c c l c c l}
\hline
$\lambda^{A}_{\mathbf{a}}$ & = &
$\max_l\{|A_{\mathbf{a}}(l)|, 1 \leq l \leq n-1\}$ &
$S^{A}_{\mathbf{a}}$
& = & $\sum_{l=1}^{n-1} |A_{\mathbf{a}}(l)|$\\
 $\lambda^{P}_{\mathbf{a}}$ &= & $\max_l
\{|P_{\mathbf{a}}(l)|, 1 \leq l \leq n-1 \}$ &
$S^{P}_{\mathbf{a}}$ &= & $\sum_{l=1}^{n-1}
|P_{\mathbf{a}}(l)|$\\
$\lambda^{A}_{\mathbf{a},\mathbf{b}}$ &= & $\max_l \{
|A_{\mathbf{a},\mathbf{b}}(l)|, \;|\;l\;| \leq n-1 \}$ &
$S^{A}_{\mathbf{a},\mathbf{b}}$& =& $\sum_{l=1-n}^{n-1}
|A_{\mathbf{a},\mathbf{b}}(l)|$\\
$\lambda^{P}_{\mathbf{a},\mathbf{b}}$ &= & $\max_l \{
|P_{\mathbf{a},\mathbf{b}}(l)|, 0 \leq l \leq n-1
\}$&$S^{P}_{\mathbf{a},\mathbf{b}}$& = &$\sum_{l=0}^{n-1}
|P_{\mathbf{a},\mathbf{b}}(l)|$\\
\hline
\end{tabular} }
\end{center}
\label{table1}
\end{table}

\subsection{Complementary sets and column correlation constraints}
A set of $m$ sequences
$\{\mathbf{a}_1,\mathbf{a}_2,...,\mathbf{a}_m \}$, each of length
$n$, is called a complementary set if
\begin{eqnarray}
\sum_{i=1}^{m} A_{\mathbf{a}_i}(l)=0 , \,\,\,\,\,\,\,\, 1 \leq l
\leq n-1 
\; .
\end{eqnarray}
When $m=2$, the binary sequence set
$\{\mathbf{a}_1,\mathbf{a}_2\}$ is called a Golay complementary
pair and $\mathbf{a}_i$, $i=1,2$, are Golay sequences. They are
known to exist for all lengths $n=2^{\alpha} 10^{\beta}
26^{\gamma}$, where $\alpha,\beta,\gamma \geq 0$~\cite{Turyn:74,
Fan}. Complementary sets
$\{\mathbf{b}_1,\mathbf{b}_2,...,\mathbf{b}_m \}$ and
$\{\mathbf{a}_1,\mathbf{a}_2,...,\mathbf{a}_m \}$ are mates if
\begin{eqnarray}
\sum_{i=1}^{m} A_{\mathbf{a}_i,\mathbf{b}_i}(l)=0 ,
\end{eqnarray}
for every $l$. The MO complementary set is a collection of
complementary sets in which any two are mates. Let
$\mathbf{M}^{k}_{m, n}$ denote the MO complementary set consisting
of ${k}$ complementary sets each having $m$ complementary
sequences of length $n$. For binary sequences, ${k}$ cannot exceed
$m$~\cite{Fan}, that is, the maximum number of mutually orthogonal
complementary sets is equal to the number of complementary
sequences in a set. Hence, $\mathbf{M}^{m}_{m, n}$ is called
\textit{a complete complementary code of order $m$}~\cite{Sue:88}.

A complementary set can be represented using a
\textit{complementary set matrix}
\begin{eqnarray}
\mathbf{C} = \left[\begin{array}{c}
\mathbf{r}_1\\
\mathbf{r}_2\\
\vdots\\
\mathbf{r}_m\\
\end{array}
\right]=\left[\begin{array}{c}
\mathbf{c}_0\\
\mathbf{c}_1\\
\vdots\\
\mathbf{c}_{n-1}\\
\end{array}
\right]^T = \left[\begin{array}{cccc}
c_{1,0} & c_{1,1}  & \cdots & c_{1,n-1}\\
c_{2,0} & c_{2,1}  & \cdots & c_{2,n-1}\\
\vdots & \vdots &  \ddots & \vdots\\
c_{m,0} & c_{m,1}  & \cdots & c_{m,n-1}\\
\end{array}
\right]_{m \times n} 
\end{eqnarray}
where $T$ denotes the matrix transpose. \textit{Row sequences},
$\mathbf{r}_i=(c_{i,0}, c_{i,1}, c_{i,2}, \cdots,c_{i,n-1})$, $1
\leq i \leq m$, are complementary sequences, that is,
$\sum_{i=1}^{m} A_{\mathbf{r}_i}(l)=0$, $1 \leq l \leq n-1$. The
main focus of this paper are properties of \textit{column
sequences}, $\mathbf{c}_j=(c_{1,j}, c_{2,j}, c_{3,j},
\cdots,c_{m,j}), \;\;0 \leq j \leq n-1$.


A set of ${k}$ mutually orthogonal complementary sets, $\{
\mathbf{C}_1, \mathbf{C}_2, ...\mathbf{C}_{k} \}$, form a MO
complementary set matrix
\begin{eqnarray}
\mathbf{M}^{k}_{m, n}=\left[\begin{array}{cccc} \mathbf{C}_1 &
\mathbf{C}_2 & \cdots & \mathbf{C}_{k}
\end{array} \right]_{m \times kn}
\end{eqnarray}
and its column sequences are denoted as $\mathbf{u}_i, 0 \leq i
\leq kn-1$. An upper bound on an autocorrelation merit of column
sequences is termed \textit{the column correlation constraint}. If
there exists at least one MO complementary set matrix satisfying a
given column correlation constraint, then this constraint is
called \textit{an achievable column correlation constraint}. For
example, let
\begin{eqnarray}
\lambda^A_{\mathbf{u}} = \max \left \{\lambda^A_{\mathbf{u}_i}, 0
\leq i \leq kn-1 \right \} ,
\end{eqnarray}
if $\lambda^A_{\mathbf{u}} \leq \lambda^A$, then
$\mathbf{M}^{k}_{m, n}$ is called a MO complementary set matrix
satisfying a column correlation constraint $\lambda^A$, and
$\lambda^A$ is an achievable column correlation constraint. We
also consider column sequences which are Golay sequences, i.e.,
Golay column sequences.

\subsection{Companion pair}

Let $\mathbf{a}$ be a sequence of length $m$, where $m$ is an even
number. We define a sequence $\mathbf{b}$ as a companion of
$\mathbf{a}$ if
\begin{eqnarray}
\mathbf{C}=\left[\begin{array}{cc} \mathbf{a} \\
\mathbf{b} \end{array} \right]^{T} \label{CompanionMatrix}
\end{eqnarray}
is a complementary set matrix which consists of $m/2$
complementary pairs. $\mathbf{C}$ is called \textit{a companion
matrix} and $(\mathbf{a}, \mathbf{b})$ is called \textit{a
companion pair}.

\begin{table} [!h]
\begin{center}\caption{Sequence Operations} {
\begin{tabular}{l c l}
\hline 
$\overleftarrow{\mathbf{a}}$ & = & $(a_{n-1}, a_{n-2}, ..., a_1, a_0)$\\
$-\mathbf{a}$ & = & $(-a_0, -a_1, ..., -a_{n-1})$\\
$\mathbf{a}^*$ &=& $(a^*_0, a^*_1, ..., a^*_{n-1})$\\
$\mathbf{a} \mathbf{b}$ & = & $(a_0, a_1, ..., a_{n-1}, b_0, b_1, ..., b_{n-1})$\\
$\mathbf{a} \!\otimes\! \mathbf{b}$ & = & $(a_0, b_0, a_1, b_1, ..., a_{n-1}, b_{n-1})$ \\
$\mathbf{a} \cdot  \mathbf{b}$ & = & $a_0b_0+a_1b_1+ ...+ a_{n-1}b_{n-1}$ \\
$f_i(\mathbf{a})$ & = & $(a_1, -a_0, a_3, -a_2, ..., a_{n-1}, -a_{n-2})$ \\
$f_c(\mathbf{a})$ & = & $(a_{\frac{n}{2}}, a_{\frac{n}{2}+1}, ...,
a_{n-1}, -a_{0}, -a_{1}
, ..., -a_{\frac{n}{2}-1} )$ \\
\hline
\end{tabular} }
\end{center}
\label{table2}
\end{table}

\subsection{Operations and extensions}
\subsubsection{Sequence and matrix operations} In Table II, we list
the following sequence operations: reversal, negation, complex
conjugation, concatenation, interleaving, and inner product.
Furthermore, we introduce two sequence reshaping functions
$f_i(\cdot)$ and $f_c(\cdot)$ defined on sequences of even length.

Let $\mathbf{C}=[c_{i,j}]$ and $\mathbf{D}$ be two matrices of
equal dimensions, then $\mathbf{C}^*=[c_{i,j}^*]$ and
$-\mathbf{C}=[-c_{i,j}]$. $\mathbf{C} \otimes \mathbf{D}$ is the
matrix whose $i$th row sequence is obtained by interleaving $i$th
row sequences of $\mathbf{C}$ and $\mathbf{D}$. $\mathbf{C}
\mathbf{D}$ denotes the matrix whose $i$th row sequence is the
concatenation of $i$th row sequences of $\mathbf{C}$ and
$\mathbf{D}$.

Let us also define a sequence set ${\cal R}^{(v)}_{\mathbf{c}}$,
which is a collection of row sequences of matrix
$\mathbf{R}_{\mathbf{c}}^{(v)}$ recursively constructed from a
sequence $\mathbf{c}$, as follows,
\begin{eqnarray} \mathbf{R}_{\mathbf{c}}^{(v)}=
\left[\begin{array}{c} \mathbf{R}_{\mathbf{c}}^{(v-1)}\mathbf{R}_{\mathbf{c}}^{(v-1)} \\
\mathbf{R}_{\mathbf{c}}^{(v-1)}(-\mathbf{R}_{\mathbf{c}}^{(v-1)})
\end{array} \right] , \;\; v=1,2,3...  \label{R_recursive}
\end{eqnarray}
where
\begin{eqnarray} \mathbf{R}_{\mathbf{c}}^{(0)}=\left[\begin{array}{c} \mathbf{c}
\\-\mathbf{c}
\end{array} \right]_{2 \times m}\label{R_0} \;.
\end{eqnarray}
Let ${\cal R}^{(v)}_{\mathbf{c},\mathbf{d}} = {\cal
R}^{(v)}_{\mathbf{c}} \cup {\cal R}^{(v)}_{\mathbf{d}}$ which
consists of $2^{v+2}$ sequences of length $2^{v}m$. For example,
${\cal R}^{(0)}_{\mathbf{c}} = \{\mathbf{c}, -\mathbf{c}\}$,
${\cal R}^{(0)}_{\mathbf{d}} = \{\mathbf{d}, -\mathbf{d}\}$, and
${\cal R}^{(0)}_{\mathbf{c},\mathbf{d}}=\{\mathbf{c},
\mathbf{d}, -\mathbf{c}, -\mathbf{d} \}$. \\

\subsubsection{Complementary set matrix extension operations} We
describe two operations for extending complementary set matrices,
namely, \textit{length-extension} and \textit{size-extension}.

\textit{Lemma 2.1}~\cite{Fan}: Let $\{\mathbf{a}, \mathbf{b}\}$ be
a complementary pair, then $\{\overleftarrow{\mathbf{b^*}},
-\overleftarrow{\mathbf{a^*}}\}$ is its mate, and both
$\{\mathbf{a} \overleftarrow{\mathbf{b^*}}, \mathbf{b}
(-\overleftarrow{\mathbf{a^*}})\}$ and $\{\mathbf{a} \otimes
\overleftarrow{\mathbf{b^*}}, \mathbf{b} \otimes
(-\overleftarrow{\mathbf{a^*}}) \}$ are complementary pairs.

$\!\!\!\!\!\!\!\!\!\!\!\!$\begin{proof} See Appendix A.\end{proof}

\textit{Lemma 2.2}: Let $\{\mathbf{a}_1,\mathbf{b}_1\}$,
$\{\mathbf{a}_2,\mathbf{b}_2\}$, ...,
$\{\mathbf{a}_m,\mathbf{b}_m\}$ be $m$ complementary pairs of
length $n$. Then, $\{\mathbf{a}_1, \mathbf{b}_1,\mathbf{a}_2,
\mathbf{b}_2, ...,\mathbf{a}_m, \mathbf{b}_m \}$ is a
complementary set of $2m$ complementary sequences.

$\!\!\!\!\!\!\!\!\!\!\!\!$\begin{proof} $\sum_{i=1}^{m} \left(
A_{\mathbf{a}_i}(l) + A_{\mathbf{b}_i}(l) \right) = 0, \; \; 1
\leq l \leq n-1 .$ \end{proof}

\textit{Lemmas 2.1-2.2} imply that, if a complementary set
consists of $m/2$ complementary pairs, the sequence length can be
recursively doubled as follows. Let
\begin{eqnarray} \mathbf{C}^{(p)} = \left[\begin{array}{c}
\mathbf{r}^{(p)}_1 \\
\mathbf{r}^{(p)}_2\\
\vdots \\
\mathbf{r}^{(p)}_{m-1}\\
\mathbf{r}^{(p)}_{m}\\
\end{array}
\right]_{m \times n^{(p)}} \;\; \textrm{and} \;\;\;\;\;\;
\mathbf{D}^{(p)} = \left[\begin{array}{c}
\overleftarrow{\mathbf{r}}^{(p)}_2\\
-\overleftarrow{\mathbf{r}}^{(p)}_1\\
\vdots \\
\overleftarrow{\mathbf{r}}^{(p)}_m\\
-\overleftarrow{\mathbf{r}}^{(p)}_{m-1}\\
\end{array}
\right]^*_{m \times n^{(p)}} \label{CD}
\end{eqnarray}
be, respectively, an $m$ by $n^{(p)}$ complementary set matrix and
its mate, where $m$ is an even number, and $\{\mathbf{r}^{(p)}_1,
\mathbf{r}^{(p)}_2 \}, \{\mathbf{r}^{(p)}_3 , \mathbf{r}^{(p)}_4
\}, ... \{ \mathbf{r}^{(p)}_{m-1}, \mathbf{r}^{(p)}_{m} \}$ are
assumed to be complementary pairs. A complementary set matrix
$\mathbf{C}^{(p+1)}$ of dimension $m$ by $n^{(p+1)}=2n^{(p)}$ can
be constructed recursively as either,
\begin{eqnarray}
\mathbf{C}^{(p+1)} &=& \mathbf{C}^{(p)} \; \mathbf{D}^{(p)}
\label{Constr1} \\
\textrm{or} \;\;\;\;\;\;\;\;\;\;\;\;\;\;\;\; \mathbf{C}^{(p+1)}
&=& \mathbf{C}^{(p)} \otimes \mathbf{D}^{(p)}. \label{Constr2}
\end{eqnarray}
We term (\ref{Constr1}) and
(\ref{Constr2})~\textit{length-extension operations}.

\textit{Lemma 2.3}~\cite{Tseng:72}: A MO complementary set matrix
$\mathbf{M}^{2k}_{2m, 2n}$ can be constructed recursively as
either,
\begin{eqnarray}
\mathbf{M}^{2{k}}_{2m, 2n}=\left[\begin{array}{rl}
\mathbf{M}^{k}_{m, n}\mathbf{M}^{k}_{m, n} , &
\!\!\!\!(-\mathbf{M}^{k}_{m, n})\mathbf{M}^{k}_{m,
n}\\
(-\mathbf{M}^{k}_{m, n})\mathbf{M}^{k}_{m, n} , & \!\!\!\!
\mathbf{M}^{k}_{m, n}\mathbf{M}^{k}_{m, n}
\end{array} \right]
\label{MO-Concatenateion}
\end{eqnarray}
or
\begin{eqnarray}
\mathbf{M}^{2{k}}_{2m, 2n}=\left[\begin{array}{rl}
\mathbf{M}^{k}_{m, n} \! \otimes \! \mathbf{M}^{k}_{m, n} , &
\!\!\!\!(-\mathbf{M}^{k}_{m, n})  \! \otimes \! \mathbf{M}^{k}_{m,
n}\\
(-\mathbf{M}^{k}_{m, n})  \! \otimes \!  \mathbf{M}^{k}_{m, n} , &
\!\!\!\! \mathbf{M}^{k}_{m, n}  \! \otimes \! \mathbf{M}^{k}_{m,
n}
\end{array} \right]
\label{MO-Interleaving}
\end{eqnarray}

$\!\!\!\!\!\!\!\!\!\!\!\!$\begin{proof} Refer to the proof of
\textit{Theorem 12-13} in~\cite{Tseng:72}. \end{proof}

We term (\ref{MO-Concatenateion}) and (\ref{MO-Interleaving})
\textit{size-extension operations}.

\section{Construction of complementary set matrices from a companion pair}

\subsection{Recursive construction}
Let ${\cal X}(m)$ denote a sequence set which consists of all
length $m$ sequences whose elements are from the alphabet ${\cal
X}$. We summarize a recursive construction of a MO complementary
set matrix $\mathbf{M}^{2^{t+1}}_{2^{t}m, 2^{t}n^{(p)}}$ with
elements from ${\cal X}$, where $n^{(p)}=2^{p+1}$, $m$ is an even
number, and $t,p=0,1,2,...$.

$\!\!\!\!\!\!$\textit{\textbf{Step 1}}: The construction starts
from a companion pair $\mathbf{c}_0$ and $\mathbf{c}_1$ which are
in ${\cal X}(m)$. They form an $m$ by $2$ companion matrix
\begin{eqnarray}
\mathbf{C}^{(0)}=\left[\begin{array}{cc} \mathbf{c}_0 \\
\mathbf{c}_1 \end{array} \right]^{T} \;\; .
\end{eqnarray}


$\!\!\!\!\!\!$\textit{\textbf{Step 2}}: By employing the
length-extension operation $p$ times, we extend $\mathbf{C}^{(0)}$
to an $m$ by $n^{(p)}=2^{p+1}$ complementary set matrix
$\mathbf{C}^{(p)}$. $\mathbf{C}^{(p)}$ and its mate
$\mathbf{D}^{(p)}$ constructed from Eq. (\ref{CD}) form a MO
complementary set matrix
\begin{eqnarray}
\mathbf{M}^{2}_{m, n^{(p)}}=\left[\begin{array}{cc}
\mathbf{C}^{(p)} & \mathbf{D}^{(p)}
\end{array} \right]_{m \times 2^{p+2}} \; .
\end{eqnarray}

$\!\!\!\!\!\!$\textit{\textbf{Step 3}}: Starting with
$\mathbf{M}^{2}_{m,n^{(p)}}$, we can construct the MO
complementary set matrix $\mathbf{M}^{2^{t+1}}_{2^{t}m,
2^{t}n^{(p)}}$ by repeating the size-extension operation $t$
times, where $p,t=0, 1, 2, ...$.

In this paper, we will alternately use either ``the constructed MO
complementary set matrix'' or, simply,
$\mathbf{M}^{2^{t+1}}_{2^{t}m, 2^{t}n^{(p)}}$ when referring to
the above constructed MO complementary set matrix.

\subsection{Companion pair design and properties}

\textit{Proposition 3.1}: Let us arrange the elements of
$\mathbf{c}_0=(c_{1,0}, c_{2,0}, ... c_{m,0})$ into $m/2$
arbitrary pairs, e.g., $(c_{x,0}, c_{y,0})$. Then, its companion
sequence $\mathbf{c}_1=(c_{1,1}, c_{2,1}, ... c_{m,1})$ can be
constructed as either
\begin{eqnarray}
&&c_{x,1}=c^*_{y,0}, \;\;\;\; c_{y,1}=-c^*_{x,0} \label{Companion1} \\
\textrm{or} \;\;\;\;\;\; &&c_{x,1}=-c^*_{y,0} , \;\;\;\;
c_{y,1}=c^*_{x,0} \label{Companion2}
\end{eqnarray}

$\!\!\!\!\!\!\!\!\!\!\!\!$\begin{proof} Let us assume
$\mathbf{c}_1$ is constructed using (\ref{Companion1}). In this
case,
\begin{eqnarray}
\mathbf{C}^{(0)}=\left[\begin{array}{c} \mathbf{c}_0 \\
\mathbf{c}_1 \end{array} \right]^{T}
=\left[\begin{array}{ccccc} ... &c_{x,0}& ...& c_{y,0}& ...  \\
...& c^*_{y,0}& ...& -c^*_{x,0}& ... \end{array} \right]^{T} \;\;
. \label{Companion}
\end{eqnarray}
Here, $\mathbf{r}^{(0)}_x = (c_{x,0}, c^*_{y,0} )$ and
$\mathbf{r}^{(0)}_y = (c_{y,0}, -c^*_{x,0} )$, respectively, the
$x$th and the $y$th row sequence of $\mathbf{C}^{(0)}$ form a
complementary pair, since
\begin{eqnarray} A_{\mathbf{r}^{(0)}_x}(l) + A_{\mathbf{r}^{(0)}_y}(l) = 0,
\,\, 1 \leq l < 2. 
\end{eqnarray}
Based on \textit{Lemma 2.2}, $\mathbf{C}^{(0)}$ is a complementary
set matrix consisting of $m/2$ complementary pairs. Hence,
$\mathbf{c}_0$ and $\mathbf{c}_1$ form a companion pair.
\end{proof}


\textit{Example 3.1}: $f_i^*(\mathbf{c}_0)$ is a companion of
$\mathbf{c}_0$, since
\begin{eqnarray}
\mathbf{C}^{(0)}=\left[\begin{array}{c} \mathbf{c}_0 \\
f^*_i(\mathbf{c}_0) \end{array} \right]^{T}
=\left[\begin{array}{ccccccc} c_{1,0} & c_{2,0}& c_{3,0} & c_{4,0}&...& c_{m-1,0}& c_{m,0}  \\
c^*_{2,0} & -c^*_{1,0}& c^*_{4,0} & -c^*_{3,0}&...& c^*_{m,0}&
-c^*_{m-1,0}
\end{array} \right]^{T} 
\end{eqnarray}
is a companion matrix. It can be verified that
$f_c^*(\mathbf{c}_0)$ is also a companion of $\mathbf{c}_0$.

The companion pair has the following properties,

\textit{Property 1 (Commutative property)}: If $\mathbf{c}_0$ is a
companion of $\mathbf{c}_1$, then $\mathbf{c}_1$ is also a
companion of $\mathbf{c}_0$.

$\!\!\!\!\!\!\!\!\!\!\!\!$\begin{proof} In
Eq.~(\ref{CompanionMatrix}), $\mathbf{C}$ is still a companion
matrix when the column sequences $\mathbf{a}$ and $\mathbf{b}$ are
switched.
\end{proof}

\textit{Property 2 (Inner product property)}: If $\mathbf{c}_0$
and $\mathbf{c}_1$ form a companion pair, then $\mathbf{c}_0 \cdot
\mathbf{c}_1 =0$.

$\!\!\!\!\!\!\!\!\!\!\!\!$\begin{proof} Based on
Eqs.~(\ref{Companion1}) and (\ref{Companion2}), $\sum_{i=1}^{m}
c_{i,0}c_{i,1}=0$.
\end{proof}

\textit{Corollary 3.1}: Binary sequences $\mathbf{c}_0$ and
$\mathbf{c}_1$  form a companion pair, if and only if
$\mathbf{c}_0 \cdot \mathbf{c}_1 =0$.

$\!\!\!\!\!\!\!\!\!\!\!\!$\begin{proof} From Property 2,
$\mathbf{c}_0 \cdot \mathbf{c}_1 =0$ is for any companion pair.
Furthermore, if $\mathbf{c}_0$ and $\mathbf{c}_1$ are binary
sequences such that $\mathbf{c}_0 \cdot \mathbf{c}_1 = 0$, there
must exist $m/2$ pairs of $(x, y)$ satisfying
$c_{x,0}c_{x,1}+c_{y,0}c_{y,1}=0$, where $1 \leq x \leq m$ and $1
\leq y \leq m$. Hence, row sequences of $\mathbf{C^{(0)}}$ can be
arranged into $m/2$ pairs, where each pair $\mathbf{r}^{(0)}_{x} =
(c_{x,0}, c_{x,1} )$ and $\mathbf{r}^{(0)}_{y} = (c_{y,0}, c_{y,1}
)$ satisfies $A_{\mathbf{r}^{(0)}_{x}}(l) +
A_{\mathbf{r}^{(0)}_{y}}(l)=0$, $1 \leq l < 2$. Consequently,
$\mathbf{C}^{(0)}$ is a companion matrix.
\end{proof}

\subsection{Column sequence properties}

\textit{Lemma 3.1}: All column sequences of the constructed
complementary set matrix $\mathbf{M}^{2}_{m,n^{(p)}} = \left[
\mathbf{C}^{(p)} \;\; \mathbf{D}^{(p)} \right ]$ are in ${\cal
R}^{(0)}_{\mathbf{c}_0,\mathbf{c}_1} = \{\pm \mathbf{c}_0, \pm
\mathbf{c}_1 \}$, where $(\mathbf{c}_0, \mathbf{c}_1)$ is the
companion pair for the construction.

$\!\!\!\!\!\!\!\!\!\!\!\!$\begin{proof} Let
\begin{eqnarray}
\mathbf{C}^{(p)}=\left[\begin{array}{c} \mathbf{c}^{(p)}_0 \\
\mathbf{c}^{(p)}_1 \\ \vdots \\ \mathbf{c}^{(p)}_{n^{(p)}-1}
\end{array} \right]^{T} . 
\end{eqnarray}
For $\mathbf{C}^{(p+1)}$ constructed by the length-extension
(\ref{Constr1}), we have,
\begin{eqnarray}
\mathbf{c}^{(p+1)}_i = \left \{ \begin {array} {cc} \mathbf{c}^{(p)}_i & 0 \leq i \leq n^{(p)}-1 \\
-\mathbf{c}^{(p)}_{i-n^{(p)}}  &
n^{(p)} \leq i \leq \frac{3}{2}n^{(p)}-1 \\
\mathbf{c}^{(p)}_{i-n^{(p)}}  & \frac{3}{2}n^{(p)} \leq i \leq
2n^{(p)}-1 \end{array} \right. \label{ColunmRelation}
\end{eqnarray}
It follows that column sequences of $\mathbf{C}^{(p+1)}$ are equal
to, or are a negation of, column sequences of $\mathbf{C}^{(p)}$.
Since column sequences of $\mathbf{C}^{(0)}$ are $\mathbf{c}_0$
and $\mathbf{c}_1$, all column sequences of $\mathbf{C}^{(p)}$,
$p=0,1,2,...$, are in ${\cal
R}^{(0)}_{\mathbf{c}_0,\mathbf{c}_1}$. In addition, for
$\mathbf{C}^{(p+1)}$ constructed using the length-extension
(\ref{Constr2}), the interleaving of the corresponding row
sequences doesn't change the column sequences and, thus, its
column sequences are also in ${\cal
R}^{(0)}_{\mathbf{c}_0,\mathbf{c}_1}$. Any column sequence of
$\mathbf{D}^{(p)}$ can be found in $\mathbf{C}^{(p+1)}$ and,
consequently, it is in ${\cal
R}^{(0)}_{\mathbf{c}_0,\mathbf{c}_1}$ as well.
\end{proof}

\textit{Lemma 3.2}: All column sequences of the constructed MO
complementary set matrix $\mathbf{M}^{2^{t+1}}_{2^{t}m,
2^{t}n^{(p)}}$ are in ${\cal
R}^{(t)}_{\mathbf{c}_0,\mathbf{c}_1}$, where $(\mathbf{c}_0,
\mathbf{c}_1)$ is the companion pair and $t, p=0, 1, 2,...$.

$\!\!\!\!\!\!\!\!\!\!\!\!$\begin{proof} Let
$\{\mathbf{u}^{(t)}_i,0 \leq i < r=2^{2t+1}n^{(p)}\}$ denote
column sequences of $\mathbf{M}^{2^{t+1}}_{2^{t}m, 2^{t}n^{(p)}}$.
The size-extension ~(\ref{MO-Concatenateion}) implies,
\begin{eqnarray}
\left \{ \begin {array} {cll}
\mathbf{u}^{(t+1)}_i  &=  -\mathbf{u}^{(t+1)}_{i+2r} & = \mathbf{u}^{(t)}_i(-\mathbf{u}^{(t)}_i)\\
\mathbf{u}^{(t+1)}_{i+r}  &=  \mathbf{u}^{(t+1)}_{i+3r} & = \mathbf{u}^{(t)}_i \mathbf{u}^{(t)}_i\\
\end{array} \right. \label{MO-ColunmRelation}
\end{eqnarray}
where $0 \leq i < r$. Based on \textit{Lemma 3.1},
$\mathbf{u}^{(0)}_i$, i.e., column sequences of
$\mathbf{M}^{2}_{m, n^{(p)}}=[ \mathbf{C}^{(p)} \;
\mathbf{D}^{(p)}]$ , are in ${\cal
R}^{(0)}_{\mathbf{c}_0,\mathbf{c}_1}$. From~(\ref{R_recursive}),
(\ref{R_0}), and (\ref{MO-ColunmRelation}), we have that
\begin{eqnarray}
\mathbf{u}^{(t)}_i \in {\cal R}^{(t)}_{\mathbf{c}_0,\mathbf{c}_1},
\;\;t=0, 1, 2,... \label{columnINk}
\end{eqnarray}
When $\mathbf{M}^{2^{t+1}}_{2^{t}m, 2^{t}n^{(p)}}$ is constructed
using size-extension~(\ref{MO-Interleaving}), the proof is
analogous.
\end{proof}



\section{Properties of the constructed complementary set matrix}

In this section, column correlation properties of the constructed
MO complementary set matrix are related to ACFs of the companion
pair. We illustrate how to satisfy a column correlation constraint
by selecting an appropriate companion pair. We also construct the
companion pair which leads to complementary set matrices with
Golay column sequences.  Since number of zeros in an
energy-normalized sequence can affect its PAPR~(see e.g.
\cite{Gavish:94,Di:03}), we also discuss the number of zeros in
column sequences at the end of this section.

\subsection{Column correlation properties}

\textit{Theorem 4.1}: MO complementary set matrix
$\mathbf{M}^{2^{t+1}}_{2^{t}m, 2^{t}n^{(p)}}$ satisfies a column
correlation constraint, if and only if the companion pair
$(\mathbf{c}_0, \mathbf{c}_1)$ is selected so that all sequences
in ${\cal R}^{(t)}_{\mathbf{c}_0,\mathbf{c}_1}$ satisfy the
constraint.

$\!\!\!\!\!\!\!\!\!\!\!\!$\begin{proof} The proof is a direct
consequence of \textit{Lemma 3.2}.  \end{proof}

\textit{Corollary 4.1}: Complementary set matrices
$\mathbf{C}^{(p)}$ and $\mathbf{D}^{(p)}$ constructed from a
companion pair $(\mathbf{c}_0, \mathbf{c}_1)$, satisfy a column
correlation constraint, if and only if $\mathbf{c}_0$ and
$\mathbf{c}_1$ satisfy the constraint.

$\!\!\!\!\!\!\!\!\!\!\!\!$\begin{proof}  The proof follows by
setting $t=0$ in \textit{Theorem 4.1}. \end{proof}

The minimum achievable column correlation constraint for the
constructed MO complementary set matrix
$\mathbf{M}^{2^{t+1}}_{2^tm,2^tn^{(p)}}$ is a function of its size
and alphabet and can be expressed as follows,
\begin{eqnarray}
\lambda_{min}(t,m) = \min \left \{ \max \left \{
\lambda_{\mathbf{c}} \; : \; \mathbf{c} \in {\cal
R}^{(t)}_{\mathbf{c}_0,\mathbf{c}_1} \right\} :  (\mathbf{c}_0,
\mathbf{c}_1)\; \textrm{is a companion pair} \;\; \textrm{and} \;
\mathbf{c}_0, \mathbf{c}_1 \in {\cal X}(m) \right \} ,
\label{MinCC}
\end{eqnarray}
where $\lambda$ is any autocorrelation merit. The following
\textit{Lemma 4.1} is a key in relating column correlation
constraints to the correlation of the companion pair. In
particular, it leads to the minimum achievable column correlation
constraint $\lambda^A_{min}(t,m)$.

\textit{Lemma 4.1}: The ACF of any column sequence
$\mathbf{u}^{(t)}_i$ of $\mathbf{M}^{2^{t+1}}_{2^tm,2^tn^{(p)}}$
can be expressed in terms of the ACF of ${\mathbf{c}_0}$ or
$\mathbf{c}_1$ , recursively, as follows,
\begin{eqnarray}
A_{\mathbf{u}_j^{(v+1)}} (l) = \left \{ \begin {array} {l}
2A_{\mathbf{u}_i^{(v)}}(l) \pm A_{\mathbf{u}_i^{(v)}}(s-l), \; 0 \leq l < s;\\
\pm A_{\mathbf{u}_i^{(v)}}(l-s),
\;\;\;\;\;\;\;\;\;\;\;\;\;\;\;\;\; s \leq l < 2s.
\end{array} \right.\label{PlusPlus_R}
\end{eqnarray}
where `$+$' holds for $j=i+r$ or $j=i+3r$, `$-$' holds for $j=i$
or $j=i+2r$, when size-extension~(\ref{MO-Concatenateion}) is
used; `$+$' holds for $j=2i+1$, `$-$' holds for $j=2i$, when
size-extension~(\ref{MO-Interleaving}) is employed;
$\mathbf{u}^{(0)}_i \in \{ \mathbf{c}_0, \mathbf{c}_1,
-\mathbf{c}_0, -\mathbf{c}_1\}$; $0 \leq i < r=2^{2v+1}n^{(p)}$,
$s=2^vm$, and $v=0,1,2,...,t-1$.

$\!\!\!\!\!\!\!\!\!\!\!\!$\begin{proof} Eq.~(\ref{PlusPlus_R}) can
be derived based on (\ref{MO-ColunmRelation}). Note that,
$\mathbf{u}^{(0)}_i \in {\cal R}^{(0)}_{\mathbf{c}_0,\mathbf{c}_1}
= \{ \mathbf{c}_0, \mathbf{c}_1,-\mathbf{c}_0, -\mathbf{c}_1 \}$
and the negation of a sequence doesn't change its ACF.\end{proof}

Similar recursive equations can be found for periodic ACFs based
on Eq.~(\ref{AtoP}).

\textit{Proposition 4.1} (\textit{A sufficient condition for
$S^A$}): Let $S^A_{\mathbf{c}_0} \leq S^A_0$, $S^A_{\mathbf{c}_1}
\leq S^A_0$, and $A_{\mathbf{c}_0}(0)=A_{\mathbf{c}_1}(0)=E$,
where $(\mathbf{c}_0, \mathbf{c}_1)$ is a companion pair. Then, a
sufficient condition for $\mathbf{M}^{2^{t+1}}_{2^tm,2^tn^{(p)}}$
to satisfy the column correlation constraint $S_t^A$ is
\begin{eqnarray}
S^A_{t} \geq 4^tS_0^A+2^{t-1}(2^t-1)E \;.\label{MOSA}
\end{eqnarray}

$\!\!\!\!\!\!\!\!\!\!\!\!$\begin{proof} Let $\{\mathbf{u}^{(t)}_i,
0 \leq i < 2^{2t+1}n^{(p)}\}$ be the column sequences of
$\mathbf{M}^{2^{t+1}}_{2^tm,2^tn^{(p)}}$. Clearly,
$\mathbf{u}^{(0)}_i \in \{ \mathbf{c}_0,
\mathbf{c}_1,-\mathbf{c}_0, -\mathbf{c}_1 \}$. Then, based on
(\ref{PlusPlus_R}), we have that
\begin{eqnarray}
S^A_{\mathbf{u}} &=& \max_{i} \left\{S^A_{\mathbf{u}^{(t)}_i}, 0
\leq i < 2^{2t+1}n^{(p)}\right\} \nonumber \\ &=&\max_{i}
\left\{\sum_{l=1}^{2^tm-1} |A_{\mathbf{u}^{(t)}_i}(l)|, 0
\leq i < 2^{2t+1}n^{(p)} \right\} \nonumber \\
&\leq& 4^tS_0^A+2^{t-1}(2^t-1)E  ,
\end{eqnarray}
for  $t=0,1,2,...$. Hence, (\ref{MOSA}) is sufficient for $S^A_t
\geq S^A_{\mathbf{u}}$ which proves the proposition.
\end{proof}

\textit{Proposition 4.2}  (\textit{A sufficient condition for
$\lambda^A$}): Let $\lambda^A_{\mathbf{c}_0} \leq \lambda^A_0$,
$\lambda^A_{\mathbf{c}_1} \leq \lambda^A_0$, and
$A_{\mathbf{c}_0}(0)=A_{\mathbf{c}_1}(0)=E$, where $(\mathbf{c}_0,
\mathbf{c}_1)$ is a companion pair. Then, a sufficient condition
for $\mathbf{M}^{2^{t+1}}_{2^tm,2^tn^{(p)}}$ to satisfy the column
correlation constraint $\lambda^A_{t}$ is
\begin{eqnarray}
\lambda^A_{t} \geq \max\left\{(2^t-1)E, \;
(2^{t+1}-1)\lambda^A_0\right\}. \label{MOlambaA}
\end{eqnarray}

$\!\!\!\!\!\!\!\!\!\!\!\!$\begin{proof} Let $\{\mathbf{u}^{(t)}_i,
0 \leq i < 2^{2t+1}n^{(p)}\}$ be the column sequences of
$\mathbf{M}^{2^{t+1}}_{2^tm,2^tn^{(p)}}$. (\ref{PlusPlus_R})
implies
\begin{eqnarray}
\lambda^A_{\mathbf{u}} &=& \max_{i}
\left\{\lambda^A_{\mathbf{u}^{(t)}_i}, 0 \leq i <
2^{2t+1}n^{(p)}\right\} \nonumber \\ &=& \max_{l,i}
\left\{|A_{\mathbf{u}^{(t)}_i}(l)|, 1 \leq l < 2^tm , 0 \leq i <
2^{2t+1}n^{(p)} \right\} \nonumber \\
&\leq& \max\left\{(2^t-1)E, (2^{t+1}-1)\lambda^A_0\right\}
\label{MO-ColunmLambda}
\end{eqnarray}
where $t=0,1,2,...$. Hence, if (\ref{MOlambaA}) holds, we have
$\lambda^A_t \geq \lambda^A_{\mathbf{u}}$.
\end{proof}

\textit{Proposition 4.3}  (\textit{A necessary condition for
$\lambda^A$}): Let $A_{\mathbf{c}_0}(0)=A_{\mathbf{c}_1}(0)=E$,
where $(\mathbf{c}_0, \mathbf{c}_1)$ is a companion pair. An
achievable column correlation constraint $\lambda^A_t$ of
$\mathbf{M}^{2^{t+1}}_{2^tm,2^tn^{(p)}}$ must satisfy
\begin{eqnarray}
\lambda^A_{t} \geq (2^t-1)E. \label{MOlambaA-LowerBound}
\end{eqnarray}

$\!\!\!\!\!\!\!\!\!\!\!\!$\begin{proof} (\ref{PlusPlus_R}) implies
that $ |A_{\mathbf{u}^{(t)}_k}(m)|=(2^t-1)E$ for
$k=2^{2t+1}n^{(p)}-1$.  Hence, $\lambda^A_t \geq \max_{l,i}
\{|A_{\mathbf{u}^{(t)}_i}(l)|, 1 \leq l < 2^tm , 0 \leq i <
2^{2t+1}n^{(p)} \} \geq (2^t-1)E$.\end{proof}

\textit{Corollary 4.2}: Let
$A_{\mathbf{c}_0}(0)=A_{\mathbf{c}_1}(0)=E$ and $\lambda^A_0 =
\max \{ \lambda^A_{\mathbf{c}_0}, \lambda^A_{\mathbf{c}_1} \}$,
where $(\mathbf{c}_0, \mathbf{c}_1)$ is a companion pair. For
$\mathbf{M}^{2^{t+1}}_{2^tm,2^tn^{(p)}}$, $t \geq 1$, when
\begin{eqnarray}
\lambda^A_{0} \leq \frac{2^t-1}{2^{t+1}-1} E,
\label{MOlambaA-LBcondition}
\end{eqnarray}
the minimum column correlation constraint
\begin{eqnarray}
\lambda^A_{min}(t,m)= (2^t-1)E \label{MOlambaA-MinimumLB}
\end{eqnarray}
is achievable.

$\!\!\!\!\!\!\!\!\!\!\!\!$\begin{proof} If
(\ref{MOlambaA-LBcondition}) holds, based on \textit{Proposition
4.2}, $\mathbf{M}^{2^{t+1}}_{2^tm,2^tn^{(p)}}$ satisfies the
column correlation constraint $\lambda^A_{t} = (2^t-1)E$. On the
other hand, \textit{Proposition 4.3} states that $\lambda^A_{t}
\geq (2^t-1)E$ must hold.
\end{proof}

\begin{figure*} [!b]
\hrulefill 
\begin{eqnarray}
\label{LongEqn} \mathbf{M}^4_{8,8} = \left[\begin{array}{c | c |
c| c} + \overline{j} - \overline{j} - j - \overline{j} & +
\overline{j} - \overline{j} - j - \overline{j} & - j + j +
\overline{j} + j & + \overline{j} - \overline{j} - j -
\overline{j}\\
j - \overline{j} - \overline{j} + \overline{j} - & j -
\overline{j} - \overline{j} + \overline{j} - & \overline{j} + j +
j - j + & j - \overline{j} - \overline{j} +
\overline{j} - \\
- \overline{j} + \overline{j} + j + \overline{j} & - \overline{j}
+ \overline{j} + j + \overline{j} & + j - j - \overline{j} - j & -
\overline{j} + \overline{j} + j + \overline{j} \\
j + \overline{j} + \overline{j} - \overline{j} + & j +
\overline{j} + \overline{j} - \overline{j} + & \overline{j} - j -
j + j - & j + \overline{j} + \overline{j} -
\overline{j} + \\
- j + j + \overline{j} + j & + \overline{j} - \overline{j} - j -
\overline{j} & + \overline{j} - \overline{j} - j - \overline{j} &
+ \overline{j} - \overline{j} - j - \overline{j} \\
\overline{j} + j + j - j + & j - \overline{j} - \overline{j} +
\overline{j} -  & j - \overline{j} - \overline{j} + \overline{j} -
& j - \overline{j} - \overline{j} + \overline{j} - \\
+ j - j - \overline{j} - j & - \overline{j} + \overline{j} + j +
\overline{j} & - \overline{j} + \overline{j} + j + \overline{j} &
- \overline{j} + \overline{j} + j + \overline{j} \\
\overline{j} - j - j + j - & j + \overline{j} + \overline{j} -
\overline{j} + & j + \overline{j} + \overline{j} - \overline{j} +
& j + \overline{j} + \overline{j} - \overline{j} + \\
\end{array} \right]_{8 \times 32}
\end{eqnarray} \label{MO-long}

\end{figure*}

\textit{Example 4.1}: To construct the complex-valued
complementary set matrices $\mathbf{C}^{(p)}$ and
$\mathbf{D}^{(p)}$ with a column correlation constraint
$\lambda^A=1$, we can choose a companion pair $\mathbf{c}_0=(+, j,
-, j)$ and $\mathbf{c}_1=f_i^{*}(\mathbf{c}_0)= (\overline{j}, -,
\overline{j}, +)$, where $+$ denotes $1$, $-$ denotes $-1$,
$j=\sqrt{-1}$ denotes the imaginary unit and $\overline{j}$
denotes $-j$, which satisfy $\lambda^A_{\mathbf{c}_i} \leq 1$,
$i=0,1$. Then, the companion matrix is
\begin{eqnarray} \textbf{C}^{(0)}
=\left[\begin{array}{cccc} + &j &- &j \\
\overline{j} &- &\overline{j} &+ \\
\end{array} \right]^{T} . 
\end{eqnarray}
By employing length-extension (\ref{Constr1}), the complementary
set matrix $\mathbf{C}^{(1)}$ and its mate $\mathbf{D}^{(1)}$ can
be obtained  as
\begin{eqnarray} \mathbf{C}^{(1)}
=\left[\begin{array}{cccc}
+ &\overline{j} &-  &\overline{j} \\
j &-  &\overline{j} &- \\
- &\overline{j} &+  &\overline{j} \\
j &+  &\overline{j} &+ \\
\end{array} \right]_{4 \times 4} , \; \mathbf{D}^{(1)}
=\left[\begin{array}{cccc}
-  & j  &-  &\overline{j} \\
\overline{j} & +  &\overline{j} &- \\
+  & j  &+  &\overline{j} \\
\overline{j} & -  &\overline{j} &+ \\
\end{array} \right]_{4 \times 4} \;.
\end{eqnarray}
Based on  \textit{Corollary 4.1}, $\mathbf{C}^{(1)}$ and
$\mathbf{D}^{(1)}$ are complementary set matrices with a column
correlation constraint $\lambda^A=1$.

\textit{Example 4.2}: Let again $\mathbf{c}_0=(+, j, -, j)$ and
$\mathbf{c}_1=f_i^{*}(\mathbf{c}_0)= (\overline{j}, -,
\overline{j}, +)$, then, any sequence $\mathbf{c}$ in ${\cal
R}^{(1)}_{\mathbf{c}_0,\mathbf{c}_1}$ satisfies
$\lambda^A_{\mathbf{c}} \leq \lambda^A=4$. Starting with
$\mathbf{M}^2_{4,4}=[\mathbf{C}^{(1)} \; \mathbf{D}^{(1)}]$
constructed in \textit{Example 4.1} and applying the
size-extension~(\ref{MO-Concatenateion}), we obtain
$\mathbf{M}^4_{8,8}$ as shown in (\ref{LongEqn}). Based on
\textit{Theorem 4.1}, $\mathbf{M}^4_{8,8}$ satisfies a column
correlation constraint $\lambda^A=4$. On the other hand,
\textit{Corollary 4.2} implies that $\mathbf{M}^4_{8,8}$ achieves
its lower bound $\lambda^A_1=E=4$, since $\lambda^A_0 =1 \leq
\frac{E}{3}$.

\textit{Example 4.3}: Let us consider how to construct
$\mathbf{M}^4_{8,8}$ satisfying a column correlation constraint
$S^A=12$. Since $t=1$ and $m=4$, based on \textit{Proposition
4.1},  a sufficient condition for $S^A_1 =12 $ is $S^A_0 \leq 2$,
for $E=4$. The companion pair $(\mathbf{c}_0, \mathbf{c}_1)$ in
\textit{Examples 4.1-4.2} satisfies $S^A_{\mathbf{c}_i} \leq
S^A_{0} = 2$, for $i=0,1$. Thus, $\mathbf{M}^4_{8,8}$ in
(\ref{LongEqn}) must also satisfy a column correlation constraint
$S^A=12$. Let $\{\mathbf{u}^{(1)}_i, 0 \leq i \leq 31\}$ denote
column sequences of $\mathbf{M}^4_{8,8}$ in (\ref{LongEqn}),  we
can verify that $\max \left\{S^A_{\mathbf{u}^{(1)}_i}, 0 \leq i
\leq 31\right \}= 12$.

\textit{Remark}: For the case $t=0$, \textit{Corollary 4.1}
implies that the correlation constraint for the companion pair is
also the column correlation constraint of
$\mathbf{M}^{2}_{m,n^{(p)}}=[\mathbf{C}^{(p)} \;
\mathbf{D}^{(p)}]$. For $t \geq 1$, based on \textit{Proposition
4.1}, small $S^A_0$ may also help in reducing the column
correlation constraint $S^A_t$. However, \textit{Corollary 4.2}
implies that it is not necessary to search for the companion pair
with smaller $\lambda^A_0$, once the lower bound
$\lambda^A_{min}(t,m)=(2^t-1)E$ has been achieved.

\subsection{Golay column sequences}

\textit{Theorem 4.2}: Column sequences of complementary set
matrices $\mathbf{C}^{(p)}$ and $\mathbf{D}^{(p)}$ are Golay
sequences, if and only if the companion sequences $\mathbf{c}_0$
and $\mathbf{c}_1$ are both Golay.

$\!\!\!\!\!\!\!\!\!\!\!\!$\begin{proof} \textit{Lemma 3.1} states
that column sequences of $\mathbf{C}^{(p)}$ and $\mathbf{D}^{(p)}$
are either $\pm \mathbf{c}_0$ or  $\pm \mathbf{c}_1$. Note that, a
negation of a Golay sequence is also a Golay sequence.
\end{proof}

We present a constructive method to obtain the companion pair from
which an $m$ by $n$ complementary set matrix with Golay column
sequences can be constructed, where $m=2^{q+1}$, $n=2^{p+1}$, $p,
q=0, 1, 2 ...$.

\textit{Theorem 4.3}: Let
\begin{eqnarray}
\mathbf{H}^{(q)}_i =  \left[\begin{array}{c}
\mathbf{h}^{(q)}_{i,0}  \\
\mathbf{h}^{(q)}_{i,1}
\end{array} \right]_{2 \times 2^q} = \left[\begin{array}{c}
\mathbf{h}^{(q-1)}_{i,0} \overleftarrow{\mathbf{h}^{(q-1)}_{i,1}}\\
\mathbf{h}^{(q-1)}_{i,1}
(-\overleftarrow{\mathbf{h}^{(q-1)}_{i,0}})
\end{array} \right] ,  \label{GolaySeed}
\end{eqnarray}
where $\mathbf{h}^{(q)}_{i,0}$ and $\mathbf{h}^{(q)}_{i,1}$ are
two row sequences of $\mathbf{H}^{(q)}_i$, $i=0,1$. The initial
matrices are
\begin{eqnarray}
\mathbf{H}^{(0)}_0 =\left[\begin{array}{cc}
+ & + \\
+ & - \\
\end{array} \right]_{2 \times 2},  \;\;
\mathbf{H}^{(0)}_1 =\left[\begin{array}{cc}
+ & - \\
+ & + \\
\end{array} \right]_{2 \times 2}. 
\end{eqnarray}
Then, $\{\mathbf{h}^{(q)}_{0,0}, \mathbf{h}^{(q)}_{0,1}\}$ and
$\{\mathbf{h}^{(q)}_{1,0},\mathbf{h}^{(q)}_{1,1}\}$ are
respectively Golay complementary pairs and, furthermore,
$f_i(\mathbf{h}^{(q)}_{0,0}) = \mathbf{h}^{(q)}_{1,0}$ and
$f_i(\mathbf{h}^{(q)}_{0,1}) = -\mathbf{h}^{(q)}_{1,1}$, for $
q=0, 1, 2 ...$.

$\!\!\!\!\!\!\!\!\!\!\!\!$\begin{proof} See Appendix C.
\end{proof}

\textit{Example 4.4}: Let $q=2$, then $\mathbf{c}_0=
\mathbf{h}^{(2)}_{0,0}=(+ + - + - - - + )$ and $\mathbf{c}_1=
\mathbf{h}^{(2)}_{1,0}=(+ - + + - + + + )$. Based on
\textit{Theorem 4.3}, $\{\mathbf{c}_0, \mathbf{h}^{(2)}_{0,1} \}$
and $\{\mathbf{c}_1, \mathbf{h}^{(2)}_{1,1} \}$ are, respectively,
Golay complementary pairs, where $\mathbf{h}^{(2)}_{0,1}=(+ - - -
- + - -)$ and $\mathbf{h}^{(2)}_{1,1}=(+ + + - - - + -) $. Thus,
the companion sequences $\mathbf{c}_0$ and $\mathbf{c}_1$ are
Golay sequences. The length-extension~(\ref{Constr2}) for $p=2$
allows for constructing the following complementary set matrix
\begin{eqnarray}
\mathbf{C}^{(2)}=\left[\begin{array}{cccccccc}
+ & - & -  & - & + & - & + & +\\
+ & - & -  & - & - & + & - & -\\
- & + & +  & + & + & - & + & +\\
+ & - & -  & - & + & - & + & +\\
- & + & +  & + & - & + & - & -\\
- & + & +  & + & + & - & + & +\\
- & + & +  & + & + & - & + & +\\
+ & - & -  & - & + & - & + & +\\
\end{array} \right]_{8\times8} 
\end{eqnarray}
whose column sequences are Golay. Hence, the PAPR of all column
sequences of $\mathbf{C}^{(2)}$ is at most two~\cite{Davis:99}.


\subsection{Number of zeros}
\textit{Proposition 4.4}: Let the companion sequence
$\mathbf{c}_0$ be a length $m$ sequence with $z$ zeros, then, any
column sequence of $\mathbf{M}^{2^{t+1}}_{2^{t}m, 2^{t}n^{(p)}}$
contains $2^tz$ zeros.

$\!\!\!\!\!\!\!\!\!\!\!\!$\begin{proof} Based on Eqs.
(\ref{Companion1}) and (\ref{Companion2}), $\mathbf{c}_0$ and its
companion $\mathbf{c}_1$ have the same number of zeros. The number
of zeros in each column sequence does not change after each
length-extension operation. Each size-extension operation doubles
the length of column sequences, as well as the number of zeros.
\end{proof}

\textit{Example 4.5}: In this example, we consider a ternary
complementary set matrix and its mate with a column correlation
constraint $S^{A}=5$. Let us set $m=8$ and $z=1$. We can find the
companion pair $\mathbf{c}_0=(+ - - + + +\, 0 \,+)$ and
$\mathbf{c}_1= f_i(\mathbf{c}_0)=(- - + + + - + \,0)$ which
satisfy $S^A_{\mathbf{c}_i} \leq 5$, $i=0,1$. The companion matrix
is
\begin{eqnarray} \textbf{C}^{(0)}
=\left[\begin{array}{c} + - - + + + \;0 + \\
- - + + + - + \;0
\end{array} \right]^{T} . 
\end{eqnarray}
Using length-extension~(\ref{Constr1}), we extend
$\mathbf{C}^{(0)}$ as
\begin{eqnarray}
\mathbf{C}^{(2)}=\left[\begin{array}{cccccccc}
+ & - & -  & - & - & + & - & -\\
- & - & +  & - & + & + & + & -\\
- & + & +  & + & + & - & + & +\\
+ & + & -  & + & - & - & - & +\\
+ & + & -  & + & - & - & - & +\\
+ & - & -  & - & - & + & - & -\\
0 & + & 0  & + & 0 & - & 0 & +\\
+ & 0 & -  & 0 & - & 0 & - & 0\\
\end{array} \right]_{8\times8} \;\; \textrm{and}\;\;\;\;
\mathbf{D}^{(2)}=\left[\begin{array}{cccccccc}
- & + & +  & + & - & + & - & -\\
+ & + & -  & + & + & + & + & -\\
+ & - & -  & - & + & - & + & +\\
- & - & +  & - & - & - & - & +\\
- & - & +  & - & - & - & - & +\\
- & + & +  & + & - & + & - & -\\
0 & - & 0  & - & 0 & - & 0 & +\\
- & 0 & +  & 0 & - & 0 & - & 0\\
\end{array} \right]_{8\times8}.
\end{eqnarray}
Based on \textit{Theorem 4.1}, complementary set matrices
$\mathbf{C}^{(2)}$ and $\mathbf{D}^{(2)}$ satisfy the column
correlation constraint $S^A=5$. Furthermore, by setting $t=0$ in
\textit{Proposition 4.4}, we have that any column sequence of
$\mathbf{C}^{(p)}$ and $\mathbf{D}^{(p)}$ has only one zero.


\section{Search for companion pairs}

\subsection{Exhaustive search algorithm}
Let ${\cal X}(n, \lambda)$ denote a subset of all sequences of
${\cal X}(n)$ which satisfy the correlation constraint $\lambda$.
For example, let ${\cal B}(2^tm) = \{ \mathbf{c} \, | \, c_i \in
{\cal B}, 1 \leq i \leq 2^tm \}$, then ${\cal B}(2^tm , S^{A}) =
\{ \mathbf{c} \, | \, \mathbf{c} \in {\cal B}(2^tm),
S_{\mathbf{c}}^A \leq S^A \}$, where ${\cal B} =\{+1 , -1\}$.
Clearly, all column sequences of binary MO complementary set
matrix $\mathbf{M}^{2^{t+1}}_{2^tm,2^tn^{(p)}}$ with a column
correlation constraint $S^A$ can be found in ${\cal B}(2^tm ,
S^{A})$. Hence, to construct a MO complementary set matrix
$\mathbf{M}^{2^{t+1}}_{2^tm,2^tn^{(p)}}$ whose column sequences
are in ${\cal X}(n, \lambda)$, we need to find a companion pair
$(\mathbf{c}_0, \mathbf{c}_1)$ such that ${\cal
R}^{(t)}_{\mathbf{c}_0,\mathbf{c}_1} \subseteq {\cal X}(n,
\lambda)$ (see \textit{Lemma 3.2}).

Let us index all $K=|{\cal X}(m)|$ sequences as $\mathbf{x}_i$,
for $1 \leq i \leq K$. When a column correlation constraint
$\lambda$ is given, desired companion pairs can be obtained by
exhaustive computer search over ${\cal X}(m)$, as described in
Table III.

\begin{table} [!h]
\begin{center}\caption{Exhaustive Search Algorithm} {
\begin{tabular}{l}
\hline j=0; \\
for $\;\;\;\; i = 1, 2, 3, ..., K$  $\;\;$ \textrm{loop} \\
$ \;\;\;\;\;$ if $\;\;\;{\cal R}^{(t)}_{\mathbf{x}_i} \subseteq
{\cal
X}(2^tm, \lambda)$, \\
$ \;\;\;\;\;\;\;\;\;\;\;$ $j=j+1$; $\mathbf{y}_j=\mathbf{x}_i$;  \\
$ \;\;\;\;\;\;\;\;\;\;\;$ for  $\;\;\;l= j-1, j-2, ... 1$, \\
$ \;\;\;\;\;\;\;\;\;\;\;\;\;\;\; \;\;\;$
check if ($\mathbf{y}_j$,$\mathbf{y}_l$) is a companion pair;\\
$ \;\;\;\;\;\;\;\;\;\;\;$ end  \\
$ \;\;\;\;\;$ end  \\
end $\;\;$ \textrm{loops}\\
\hline
\end{tabular} }
\end{center}
\label{table3}
\end{table}

Note that the ACFs of sequences in ${\cal R}^{(t)}_{\mathbf{x}_i}$
can be computed recursively using (\ref{PlusPlus_R}). For binary
sequences, we can simply check if $\mathbf{x} \cdot \mathbf{y} =
0$ to determine the companion pair.



\subsection{Minimum achievable column correlation constraint}

The exhaustive search algorithm in Table III can be easily
modified to search for the companion pair with a minimum
achievable column correlation constraint. However, the computing
load is heavy, especially for large $m$ and $t$. Let $t=0$ in
(\ref{MinCC}), the companion pair for the construction of
$\mathbf{M}^{2}_{m,n^{(p)}} = [\mathbf{C}^{(p)} \;
\mathbf{D}^{(p)}]$ with a minimum achievable column correlation
constraint is
\begin{eqnarray}
(\mathbf{c}_0, \mathbf{c}_1) = \arg \min_{(\mathbf{x},\mathbf{y})}
\left \{ \max \left \{ \lambda_{\mathbf{x}},\lambda_{\mathbf{y}}
\right \} : (\mathbf{x}, \mathbf{y})\; \textrm{is a companion
pair} \; \textrm{and} \; \mathbf{x}, \mathbf{y} \in {\cal X}(m)
\right \} .
\end{eqnarray}
Based on \textit{Propositions 4.1-4.3}, the above companion pair
may also lead to the MO complementary set matrix
$\mathbf{M}^{2^{t+1}}_{2^tm,2^tn^{(p)}}$ with a reduced column
correlation constraint for $t \geq 1$. Hence, in the following, we
consider companion pairs leading to an achievable or a minimum
achievable column correlation constraint
$\lambda_{min}(m)=\lambda_{min}(t=0,m)$ for the case $t=0$ only.
Table~IV lists binary companion pairs with minimum achievable
column correlation constraints $\lambda^A_{min}(m)$ and
$S^A_{min}(m)$. It can be observed that most of these companion
pairs can achieve $\lambda^A_{min}(m)$ and $S^A_{min}(m)$
simultaneously.

\begin{table} [!h]
\begin{center}\caption{Binary Companion Pairs} {
\begin{tabular}{| c | c | c |l || c| c| c| l|}
\hline $m$ & $\lambda^A_{min}$ &$S^A_{min}$& companion pair&$m$&$\lambda^A_{min}$&$S^A_{min}$& companion pair\\
\hline 2 & 1 &  1 &  $\begin{array}{c} -+\\ ++ \end{array}$&
       12 & 2 & 8& $\begin{array}{c} ++++-+-++--+\\--++-----+-+\end{array}$ \\
 \hline 4 & 1 &  2& $\begin{array}{c} ---+\\ -+++ \end{array}$ &
       14 & 2 & 13 &  $\begin{array}{c} +-+------++--+\\----++--++-+-+ \end{array}$\\
 \hline 6 & 2 & 5 & $\begin{array}{c} -+---+\\ --+-++ \end{array}$
         & 16 & 2 &  $\diagup$ & $\begin{array}{c} -+++------+--+-+\\ +--++-+-+------+ \end{array}$\\
 \hline 8 & 2 & 6& $\begin{array}{c} -+----++\\+++--+-+\end{array}$
        & 16 &  $\diagup$ & 12& $\begin{array}{c} ++--+++++-+-+--+\\-+++-+++-+--+-++ \end{array}$ \\
 \hline 10 & 2 & 9& $\begin{array}{c}   +-+++++--+\\---++--+-+
\end{array}$
      & 18 & 2 & 17 &  $\begin{array}{c} ++++----+-+--++--+\\-+-++-+-----++--++ \end{array}$\\
\hline
\end{tabular} }
\end{center}
\label{TableBinarySeed}
\end{table}

When $m$ is large, the exhaustive computer search is infeasible.
Hence, the existence of a companion pair satisfying a correlation
constraint is an important problem considered in the next section.

\section{The existence of companion pairs}
The existence of a sequence of a desired correlation constraint
has been studied in literature. For example, binary sequences with
$\lambda^A=1$ exist only for lengths $2,3,4,5,7,11$ and $13$, and
are called binary Barker sequences; binary m-sequences~\cite{Fan}
with $\lambda^P=1$ exist for length $m=2^{l}-1$, $l=2,3,4...$. In
this section, we exploit the correlation properties of companion
pairs and analyze their existence for correlation constraints
$\lambda^{A}$ and $\lambda^{P}$.

\subsection{Correlation properties of the companion pair constructed using two arbitrary sequences}
In the following Cases 1 and 2 we illustrate how a companion pair
of length $m$ can be formed using two arbitrary sequences
$\mathbf{s}_0$ and $\mathbf{s}_1$ of length $m/2$. We study the
correlation properties of the companion pair
$(\mathbf{c}_0,\mathbf{c}_1)$ constructed from $\mathbf{s}_0$ and
$\mathbf{s}_1$.

\textbf{\textit{Case 1:}} Let $\mathbf{c}_0 = \mathbf{s}_0 \otimes
\mathbf{s}_1$ and $\mathbf{c}_1=\mathbf{s}^*_1 \otimes
(-\mathbf{s}^*_0)$. Then $\mathbf{c}_1$ is a companion of
$\mathbf{c}_0$ since $\mathbf{c}_1 = f_i^{*}(\mathbf{c}_0)$.

In this case, ACFs of the companion pair can be expressed in terms
of the ACFs of $\mathbf{s}_0$ and $\mathbf{s}_1$ and their
crosscorrelation functions,
\begin{eqnarray}
A_{\mathbf{c}_0} (l) \!\!\!\!&=&\!\!\!\! \left \{ \begin {array}
{ll} A_{\mathbf{s}_0,\mathbf{s}_1}(\frac{l-1}{2}) +
A_{\mathbf{s}_0,\mathbf{s}_1}(\frac{-l-1}{2}), & l\in odd\\
A_{\mathbf{s}_0}(\frac{l}{2}) + A_{\mathbf{s}_1}(\frac{l}{2}),  &
l \in even
\end{array} \right. \label{Scor1} \\
A_{\mathbf{c}^{*}_1} (l) \!\!\!\!&=&\!\!\!\! \left \{ \begin
{array} {ll} -A_{\mathbf{s}_0,\mathbf{s}_1}(\frac{l+1}{2}) -
A_{\mathbf{s}_0,\mathbf{s}_1}(\frac{-l+1}{2}),  & \! \! l\in odd\\
A_{\mathbf{s}_0}(\frac{l}{2}) + A_{\mathbf{s}_1}(\frac{l}{2}), &
\! \! \! \! l \in even
\end{array} \right. \label{Scor2}
\end{eqnarray}
\begin{eqnarray}
P_{\mathbf{c}_0} (l) \!\!\!\!&=&\!\!\!\! \left \{ \begin {array}
{ll} P_{\mathbf{s}_0,\mathbf{s}_1}(\frac{l-1}{2}) +
P_{\mathbf{s}_0,\mathbf{s}_1}(\frac{m-l-1}{2}), & \!\!  l\in odd\\
P_{\mathbf{s}_0}(\frac{l}{2}) +
P_{\mathbf{s}_1}(\frac{l}{2}),  & \!\! l \in even\\
\end{array} \right. \label{Scor3} \\
P_{\mathbf{c}^*_1} (l) \!\!\!\!&=&\!\!\!\! \left \{ \begin {array}
{ll} -P_{\mathbf{s}_0,\mathbf{s}_1}(\frac{l+1}{2}) -
P_{\mathbf{s}_0,\mathbf{s}_1}(\frac{m-l+1}{2}), & \!\!\!\!\!  l\in odd\\
P_{\mathbf{s}_0}(\frac{l}{2}) +
P_{\mathbf{s}_1}(\frac{l}{2}), & \!\!\!\!\!  l \in even\\
\end{array} \right. \label{Scor4}
\end{eqnarray}
where $ 0 \leq l \leq m-1$.

\textit{Lemma 6.1}: Let $\mathbf{c}_0 = \mathbf{s}_0 \otimes
\mathbf{s}_1$ and $\mathbf{c}^{*}_1 = \mathbf{s}_1 \otimes
(-\mathbf{s}_0)$, then
\begin{eqnarray}
\left \{ \begin {array} {c} \lambda^{A}_{\mathbf{c}_i} \leq \max
\{\lambda^{A}_{\mathbf{s}_0}+\lambda^{A}_{\mathbf{s}_1},
2\lambda^{A}_{\mathbf{s}_0,\mathbf{s}_1}\}  \\
\lambda^{P}_{\mathbf{c}_i} \leq \max
\{\lambda^{P}_{\mathbf{s}_0}+\lambda^{P}_{\mathbf{s}_1},
2\lambda^{P}_{\mathbf{s}_0,\mathbf{s}_1}\} \end{array}
\right.\label{Sconstr1}
\end{eqnarray}
and
\begin{eqnarray}
\left \{ \begin {array} {c} S^{A}_{\mathbf{c}_i} \leq
S^{A}_{\mathbf{s}_0} + S^{A}_{\mathbf{s}_1} +
S^{A}_{\mathbf{s}_0,\mathbf{s}_1} \\ S^{P}_{\mathbf{c}_i} \leq
S^{P}_{\mathbf{s}_0} + S^{P}_{\mathbf{s}_1} +
2S^{P}_{\mathbf{s}_0,\mathbf{s}_1}
\end{array}
\right. \label{Sconstr2}
\end{eqnarray} where $i=0,1$.

$\!\!\!\!\!\!\!\!\!\!\!\!$\begin{proof} See Appendix C.
\end{proof}

\textbf{\textit{Case 2:}} Let $\mathbf{c}_0 = \mathbf{s}_0
\mathbf{s}_1$ and $\mathbf{c}_1=\mathbf{s}^*_1(-\mathbf{s}^*_0)$.
Then $\mathbf{c}_1$ is a companion of $\mathbf{c}_0$ since
$\mathbf{c}_1 = f_c^{*}(\mathbf{c}_0)$.

The aperiodic ACFs of the companion pair can be expressed as
\begin{eqnarray}
A_{\mathbf{c}_0} (l) \!\!\!\!&=&\!\!\!\! \left \{ \begin {array}
{ll} A_{\mathbf{s}_0}(l)+A_{\mathbf{s}_1}(l)+
A_{\mathbf{s}_0,\mathbf{s}_1}(l-\frac{m}{2}), & 0 \leq l < \frac{m}{2} \\
A_{\mathbf{s}_0,\mathbf{s}_1}(l-\frac{m}{2}) & \frac{m}{2} \leq l
< m
\end{array} \right. \label{ConScor1} \\
A_{\mathbf{c}_1} (l) \!\!\!\!&=&\!\!\!\! \left \{ \begin {array}
{ll} A_{\mathbf{s}_0}(l)+A_{\mathbf{s}_1}(l)-
A_{\mathbf{s}_0,\mathbf{s}_1}(\frac{m}{2}-l), & 0 \leq l < \frac{m}{2} \\
-A_{\mathbf{s}_0,\mathbf{s}_1}(\frac{m}{2}-l) & \frac{m}{2} \leq l
< m
\end{array} \right. \label{ConScor2}
\end{eqnarray}

\textit{Lemma 6.2}: Let $\mathbf{c}_0 = \mathbf{s}_0 \mathbf{s}_1$
and $\mathbf{c}^{*}_1 = \mathbf{s}_1 (-\mathbf{s}_0)$, then
\begin{eqnarray}
\begin {array} {c} \lambda^{A}_{\mathbf{c}_i} \leq
\lambda^{A}_{\mathbf{s}_0}+\lambda^{A}_{\mathbf{s}_1}+
\lambda^{A}_{\mathbf{s}_0,\mathbf{s}_1}, \;\;\;\; i=0,1 .
\end{array} \label{BadLambda}
\end{eqnarray}
$\!\!\!\!\!\!\!\!\!\!\!\!$\begin{proof} The proof is along the
lines of the proof of \textit{Lemma 6.1}.
\end{proof}

\subsection{Existence}
Without loss of generality, we assume that $\mathbf{s}_i$ are
complex-valued sequences of length $m/2$ and
$A_{\mathbf{s}_i}(0)=P_{\mathbf{s}_i}(0)=m/2$, $i=0,1$.

\textit{Lemma 6.3} (Welch bound~\cite{Welch:74}): Let
$\{\mathbf{s}_i, i=0,1,...,K-1 \}$, denote a set of $K$
complex-valued sequences of length $N$. If
$A_{\mathbf{s}_i}(0)=P_{\mathbf{s}_i}(0)=N$ for all $i$, then,
\begin{eqnarray}
P_{max} &\geq& N \sqrt{\frac{K-1}{NK-1}} \label{WelchP} \\
A_{max} &\geq& N\sqrt{\frac{K-1}{2NK-K-1}} \label{WelchA}
\end{eqnarray}
where
\begin{eqnarray}
P_{max} &=& \max_{0 \leq i,j < K, i \neq j} \{
\lambda^{P}_{\mathbf{s}_i}, \lambda^{P}_{\mathbf{s}_i, \mathbf{s}_j} \} \;  \\
A_{max} &=& \max_{0 \leq i,j < K, i \neq j} \{
\lambda^{A}_{\mathbf{s}_i}, \lambda^{A}_{\mathbf{s}_i,
\mathbf{s}_j} \} .
\end{eqnarray}

$\!\!\!\!\!\!\!\!\!\!\!\!$\begin{proof} The proof can be found in
~\cite{Welch:74}. \end{proof}

\subsubsection{Column correlation constraint $\lambda^A$} The
following \textit{Theorems 6.1-6.2} restate the companion pair
existence conditions from \textit{Theorems 4.1-4.2} in terms of
the $\{ \mathbf{s}_0, \mathbf{s}_1\}$ pair existence conditions
from \textit{Lemmas 6.1-6.2}.

\textit{Theorem 6.1}: MO complementary set matrix
$\mathbf{M}^{2}_{m,n^{(p)}}$ with a column correlation constraint
$\lambda^{A}$ exists if there exists a sequence pair $\{
\mathbf{s}_0, \mathbf{s}_1\}$ with
$A_{max}=\frac{1}{2}\lambda^{A}$.

$\!\!\!\!\!\!\!\!\!\!\!\!$\begin{proof} \textit{Theorem 4.1}
states that MO complementary set matrix
$\mathbf{M}^{2}_{m,n^{(p)}}$ satisfying the column correlation
constraint $\lambda^{A}$ exists, if and only if we can find a
companion pair $(\mathbf{c}_0, \mathbf{c}_1)$, such that,
\begin{eqnarray} \lambda^{A}_{\mathbf{c}_i}
\leq  \lambda^{A}, \,\,\,\, i=0,1 \label{suffi_1}
\end{eqnarray}
Based on (\ref{Sconstr1}), a sufficient condition for
(\ref{suffi_1}) is,
\begin{eqnarray} \lambda^{A}_{\mathbf{s}_i}
\leq \frac{\lambda^{A}}{2}, \,\,i=0,1\,\,\,\, \textrm{and}
\,\,\,\, \lambda^{A}_{\mathbf{s}_0,\mathbf{s}_1} \leq
\frac{\lambda^{A}}{2} \label{ExistenceGood}
\end{eqnarray}
Based on (\ref{BadLambda}),
\begin{eqnarray} \lambda^{A}_{\mathbf{s}_i}
\leq \frac{\lambda^{A}}{3}, \,\,i=0,1\,\,\,\, \textrm{and}
\,\,\,\, \lambda^{A}_{\mathbf{s}_0,\mathbf{s}_1} \leq
\frac{\lambda^{A}}{3} \label{ExistenceBad}
\end{eqnarray}
Hence, by comparing (\ref{ExistenceGood}) and
(\ref{ExistenceBad}), we can set $A_{max} =
\frac{1}{2}\lambda^{A}$.\end{proof}

\textit{Proposition 6.1}: Sequence pair $\{ \mathbf{s}_0,
\mathbf{s}_1\}$ of length $\frac{m}{2}$ with
$A_{max}=\frac{1}{2}\lambda^{A}$ exists only if
\begin{eqnarray}
\lambda^{A} \geq \frac{m}{\sqrt{2m-3}} \label{Theorem62}
\end{eqnarray}

$\!\!\!\!\!\!\!\!\!\!\!\!$\begin{proof}Let
$A_{max}=\frac{1}{2}\lambda^{A}$, $K=2$, and $N=\frac{m}{2}$ in
(\ref{WelchA}) of \textit{Lemma 6.3}, then (\ref{Theorem62})
follows.\end{proof}

\textit{Corollary 6.1}: Let $\mathbf{c}_0 = \mathbf{s}_0 \otimes
\mathbf{s}_1$ and $\mathbf{c}_1=\mathbf{s}_1 \otimes
(-\mathbf{s}_0)$ be a binary companion pair of length $m$, and
$\mathbf{u}_i$ denote column sequences of the constructed
$\mathbf{M}^{2}_{m,n^{(p)}}$, $0 \leq i < 2n^{(p)}$. Then,
$\lambda_W^{A} \leq \lambda^A_{\mathbf{u}} \leq \lambda_B^{A}$,
where
\begin{eqnarray}
\lambda^A_{\mathbf{u}} &=& \max_{i}
\left \{\lambda^A_{\mathbf{u}^{(t)}_i}, 0 \leq i < 2n^{(p)} \right \} \\
\lambda_B^{A} &=& \max \left
\{\lambda^{A}_{\mathbf{s}_0}+\lambda^{A}_{\mathbf{s}_1},
2\lambda^{A}_{\mathbf{s}_0,\mathbf{s}_1}\right \}, \\
\lambda_W^{A} &=& \lceil \frac{m}{\sqrt{2m-3}} \rceil.
\end{eqnarray}

$\!\!\!\!\!\!\!\!\!\!\!\!$\begin{proof} $ \lambda_W^{A}$ is
derived from \textit{Theorem 6.1} and (\ref{Theorem62}) by noting
that $\lambda^A_{\mathbf{u}}$ is an integer for binary sequences.
$\lambda_B^{A}$ follows from \textit{Lemma 6.1}.\end{proof}

\subsubsection{Column correlation constraint} $\lambda^P$

\textit{Theorem 6.2}: MO complementary set matrix
$\mathbf{M}^{2}_{m,n^{(p)}}$ with a column correlation constraint
$\lambda^{P}$ exists, if there exists $\{\mathbf{s}_0,
\mathbf{s}_1 \}$ of length $m/2$ with
$P_{max}=\frac{1}{2}\lambda^{P}$.

$\!\!\!\!\!\!\!\!\!\!\!\!$\begin{proof} The proof follows along
the lines of the proof of \textit{Theorem 6.1} and is
omitted.\end{proof}

\textit{Proposition 6.2}: A length $m/2$ sequence pair
$\{\mathbf{s}_0, \mathbf{s}_1 \}$ with
$P_{max}=\frac{1}{2}\lambda^{P}$ exists only if,
\begin{eqnarray}
\lambda^{P} \geq \frac{m}{\sqrt{m-1}} \label{Theorem64}
\end{eqnarray}

$\!\!\!\!\!\!\!\!\!\!\!\!$\begin{proof} Setting
$P_{max}=\frac{1}{2}\lambda^{P}$, $K=2$ and $N=\frac{m}{2}$ in
(\ref{WelchP}) leads to (\ref{Theorem64}).
\end{proof}

\subsection{Achievable column correlation constraints}

\textit{Theorems 6.1-6.2} suggest searching for sequences
$\mathbf{s}_0$ and $\mathbf{s}_1$ of length $m/2$ with good
autocorrelation and crosscorrelation merits to form a companion
pair with a small achievable column correlation constraint. Former
is a long standing problem (e.g. see \cite{Somaini:75, Sarwate:79,
Hai:96, Mow:97}). In~\cite{Hai:96}, good binary sequence pairs
with small $\lambda^{A}_{\mathbf{s}_0}$,
$\lambda^{A}_{\mathbf{s}_1}$ and
$\lambda^{A}_{\mathbf{s}_0,\mathbf{s}_1}$ were found by using
simulated annealing search algorithm, and were listed in Tables I
and II. Based on \textit{Corollary 6.1}, we present
$\lambda_W^{A}$ and $\lambda_B^{A}$ of their corresponding binary
companion pairs in Table V, where the reference~\cite{Hai:96}
indicates that data is obtained by using sequences from this
reference. However, the cost function for the simulated annealing
in~\cite{Hai:96} is not optimal in our case. We instead minimize
the cost function
\begin{eqnarray}
f(\mathbf{s}_0,\mathbf{s}_1)=\max \left
\{\lambda^{A}_{\mathbf{s}_0}+\lambda^{A}_{\mathbf{s}_1},
2\lambda^{A}_{\mathbf{s}_0,\mathbf{s}_1}\right\}
\label{OurCostFunction}
\end{eqnarray}
to obtain an improved $\lambda_B^A$.

In Table VI, sequence pairs $\{\mathbf{s}_0, \mathbf{s}_1\}$ of
length $m/2= 63, 84$ and $100$ obtained using simulated annealing
based on~(\ref{OurCostFunction}) are presented. The corresponding
ACF merit $\lambda^A_{\mathbf{u}}$ is calculated and compared to
that of the sequence pairs from~\cite{Hai:96}. The proposed
sequence pairs lead to companion pairs with an improved
autocorrelation correlation merit.
\begin{table} [!h]
\begin{center}\caption{$\lambda_W^{A}$ and $\lambda_B^{A}$ for Long Binary Seed Sequences} {
\begin{tabular}{ l || c | c |c | c| c| c| c| c| c| c| c| c| c| c| c| c}
\hline $m$ &  62 & 74& 82&106 & 118 & 122 &126 & 134 &146 &158 &168 &182  &186 &200 &218 &240\\
\hline 
$\lambda_W^{A}$ & 6 & 7 & 7 & 8 & 8 & 8 & 8 &
9&  9& 9 & 10 & 10 &  10 & 11&  11 &   11\\
\hline $ \lambda_B^{A}$~\cite{Hai:96}
&  16 & 18& 16& 18 &  20  & 22  & 22  &22  &24 & 24  & 28  &24  &24  &28  &30  &28\\
\hline $\lambda_B^{A}$
 &  13 & 15& 15& 18 &  18  & 18  & 19  &20  &22 & 22  & 24  &24  &24  &27  &28  &28\\
\hline
\end{tabular} }
\end{center}
\label{TableBinaryLongSeed}
\end{table}


\begin{table} [!b]
\begin{center}\caption{Achievable $\lambda^{A}$ for Long Binary Seed Sequences} {
\begin{tabular}{ c ||  c| c| c}
\hline $m$ &  merits & $\mathbf{s}_0$ (or $\mathbf{s}_1$ ) & $\mathbf{s}_1$ (or $\mathbf{s}_0$)\\
\hline $126$ & $\begin{array}{c} \lambda_B^A=19\\\lambda^A_{\mathbf{u}}=17\\
\lambda_B^A=22 \;[20] \\
\lambda^A_{\mathbf{u}}=17 \; [20]
\end{array}$
&$\begin{array}{c}
 +-+-+--+-+-+++++++--+\\--+-+++--+---+++---++\\--++-++----++-+---+++
\end{array}$
&$\begin{array}{c}
 ++-+++-++--++----+-++\\-+++--++++--+++++-+-+\\+++-+----++-+-+-+---+
\end{array}$\\
\hline $168$ &  $\begin{array}{c}
\lambda_B^A=24\\\lambda^A_{\mathbf{u}}=20 \\ \lambda_B^A=28\;[20] \\
\lambda^A_{\mathbf{u}}=21\;[20] \end{array}$ &$\begin{array}{c}
 +--+--+++++----+-++--\\+-++--++--++-++----++\\-+++-+-+---+++-++++--\\+-++-+-++------+-----
\end{array}$ &$\begin{array}{c}
+---+------++---++++-\\--+---++-+++---+-+-++\\-+-+--+++++---++++++-\\++-++-+++-+-++-+--+--
\end{array}$\\
\hline $200$ &  $\begin{array}{c}
\lambda_{B}^A=27\\\lambda^A_{\mathbf{u}}=23\\ \lambda_B^A=28 \;[20] \\
\lambda^A_{\mathbf{u}}=25 \;[20]
\end{array}$
&$\begin{array}{c}
-+++++-+++++----+-++\\++---++-+-+-++--+-++\\++++---++++-+-+--+-+\\-+++++-++---+-++--+-\\-+--++++---+-+-+-+++
\end{array}$
&$\begin{array}{c}
--+++-++-+---+++--++\\+-++--+-+--+--+-+---\\-----++--+----+++--+\\-+++++----++--+-+++-\\-++-+-++-++-+-+++--+
\end{array}$\\
\hline
\end{tabular} }
\end{center}
\label{TableBinaryLongSeed}
\end{table}

\section{Conclusion}

We have considered a construction algorithm for MO complementary
set matrices satisfying a column correlation constraint. The
algorithm recursively constructs the MO complementary set matrix,
starting from a companion pair. We relate correlation properties
of column sequences to that of the companion pair and illustrate
how to select an appropriate companion pair to satisfy a given
column correlation constraint.  We also reveal a method to
construct the Golay companion pair which leads to the
complementary set matrix with Golay column sequences. An
exhaustive computer search algorithm is described which helps in
searching for companion pairs with a minimum achievable column
correlation constraint. Exhaustive search is infeasible for
relatively long sequences. Hence, we instead suggest a strategy
for finding companion pairs with a small, if not minimum, column
correlation constraint. Based on properties of the companion pair,
the strategy suggests a search for any two shorter sequences by
minimizing a cost function in terms of their autocorrelation and
crosscorrelation merits, from which the desired companion pair can
be formed. An improved  cost function is derived to further reduce
the achievable column correlation constraint $\lambda^A$. By
exploiting the well-known Welch bound, sufficient conditions for
the existence of companion pairs are also derived for column
correlation constraints $\lambda^A$ and $\lambda^P$.

We have left the general problem of finding MO complementary set
matrices with a minimum column correlation constraint as an open
question. An important step towards solving the general problem is
to find new construction approaches for MO complementary set
matrices. A design algorithm based on N-shift cross-orthogonal
sequences can be found in [24]. However, their column correlation
properties are intractable.

\begin{appendix}
\subsection{Proof of Lemma 2.1}

Let us first prove that $\{ \overleftarrow{\mathbf{b^*}},
-\overleftarrow{\mathbf{a^*}}\}$ is a mate of $\{\mathbf{a},
\mathbf{b}\}$. A proof for binary sequences can be found
in~\textit{Theorem 11} of~\cite{Tseng:72}. For complex-valued
sequences, the complementarity of $\{
\overleftarrow{\mathbf{b^*}}, -\overleftarrow{\mathbf{a^*}}\}$
follows from
\begin{eqnarray}
A_{\overleftarrow{\mathbf{b^*}}}(l) +
A_{-\overleftarrow{\mathbf{a^*}}}(l) &=&
(A_{\overleftarrow{\mathbf{b}}}(l))^* +
(A_{-\overleftarrow{\mathbf{a}}}(l))^* \nonumber \\
&=& A_{\mathbf{b}}(l) + A_{-\mathbf{a}}(l) \nonumber \\
&=& A_{\mathbf{b}}(l) + A_{\mathbf{a}}(l) \nonumber \\
&=& 0 \nonumber
\end{eqnarray}
for $1 \leq l \leq n-1$, where $n$ denotes the sequence length. We
further show that the pair $\{ \overleftarrow{\mathbf{b^*}},
-\overleftarrow{\mathbf{a^*}}\}$ is orthogonal to $\{ \mathbf{a},
\mathbf{b}\}$ in the complementary sense, as follows
\begin{eqnarray}
A_{\mathbf{a},\overleftarrow{\mathbf{b^*}}}(l) +
A_{\mathbf{b},-\overleftarrow{\mathbf{a^*}}}(l) &=&
A_{\mathbf{a},\overleftarrow{\mathbf{b^*}}}(l) -
A_{\mathbf{b},\overleftarrow{\mathbf{a^*}}}(l) \nonumber \\
&=& (A_{\mathbf{a^*},\overleftarrow{\mathbf{b}}}(l))^* -
A_{\mathbf{b},\overleftarrow{\mathbf{a^*}}}(l) \nonumber\\
&=& A_{\mathbf{b},\overleftarrow{\mathbf{a^*}}}(l) -
A_{\mathbf{b},\overleftarrow{\mathbf{a^*}}}(l) \nonumber \\
&=& 0 \nonumber
\end{eqnarray}
for every $l$.

Refer to the proofs of \textit{Theorem 6} and \textit{Theorem 13}
from~\cite{Tseng:72}. If $\{\mathbf{a}_1, \mathbf{b}_1\}$ is a
complementary pair and $\{\mathbf{a}_2, \mathbf{b}_2 \}$ is one of
its mates, then both $\{\mathbf{a}_1\mathbf{a}_2,
\mathbf{b}_1\mathbf{b}_2 \}$ and $\{\mathbf{a}_1 \otimes
\mathbf{a}_2, \mathbf{b}_1 \otimes \mathbf{b}_2 \}$ are
complementary pairs. This completes the proof of~\textit{Lemma
2.1}.

\subsection{Proof of Theorem 4.3}
\begin{eqnarray}
\mathbf{H}^{(0)}_0 = \left[\begin{array}{c}
\mathbf{h}^{(0)}_{0,0}  \\
\mathbf{h}^{(0)}_{0,1}
\end{array} \right]= \left[\begin{array}{cc}
+ & + \\
+ & -
\end{array} \right]_{2 \times 2} \nonumber
\end{eqnarray}
and
\begin{eqnarray}
\mathbf{H}^{(0)}_1 = \left[\begin{array}{c}
\mathbf{h}^{(0)}_{1,0}  \\
\mathbf{h}^{(0)}_{1,1}
\end{array} \right]= \left[\begin{array}{cc}
+ & - \\
+ & +
\end{array} \right]_{2 \times 2} \nonumber
\end{eqnarray}

It can be verified that
$\{\mathbf{h}^{(0)}_{0,0},\mathbf{h}^{(0)}_{0,1}\}$ and
$\{\mathbf{h}^{(0)}_{1,0},\mathbf{h}^{(0)}_{1,1}\}$ are
respectively Golay complementary pairs. Based on \textit{Lemma
2.1}, $\{\mathbf{h}^{(q)}_{0,0},\mathbf{h}^{(q)}_{0,1}\}$ and
$\{\mathbf{h}^{(q)}_{1,0},\mathbf{h}^{(q)}_{1,1}\}$, $q=1,2,3...$,
constructed from (\ref{GolaySeed}) are guaranteed to be Golay
complementary pairs.

We observe that
$f_i(\mathbf{h}^{(0)}_{0,0})=\mathbf{h}^{(0)}_{1,0}$ and
$f_i(\mathbf{h}^{(0)}_{0,1})=-\mathbf{h}^{(0)}_{1,1}$. Let
$f_i(\mathbf{h}^{(q)}_{0,0})=\mathbf{h}^{(q)}_{1,0}$,
$f_i(\mathbf{h}^{(q)}_{0,1})=-\mathbf{h}^{(q)}_{1,1}$, then,
\begin{eqnarray}
f_i(\mathbf{h}^{(q+1)}_{0,0}) &=& f_i(\mathbf{h}^{(q)}_{0,0} \,\,
\overleftarrow{\mathbf{h}^{(q)}_{0,1}}) =
f_i(\mathbf{h}^{(q)}_{0,0}) \,\,
\overleftarrow{(-f_i(\mathbf{h}^{(q)}_{0,1}))}=
\mathbf{h}^{(q)}_{1,0} \,\, \overleftarrow{\mathbf{h}^{(q)}_{1,1}}
= \mathbf{h}^{(q+1)}_{1,0} \nonumber \;.
\end{eqnarray}
In a similar way, we have that $f_i(\mathbf{h}^{(q+1)}_{0,1})= -
\mathbf{h}^{(q+1)}_{1,1}$. This ends the proof.

\subsection{Proof of Lemma 6.1}
We give the proof for $ \lambda^{P}_{\mathbf{c}_0} \leq \max
\left\{\lambda^{P}_{\mathbf{s}_0}+\lambda^{P}_{\mathbf{s}_1},
2\lambda^{P}_{\mathbf{s}_0,\mathbf{s}_1}\right\} $ and $
S^{A}_{\mathbf{c}_0} \leq S^{A}_{\mathbf{s}_0} +
S^{A}_{\mathbf{s}_1} + S^{A}_{\mathbf{s}_0,\mathbf{s}_1} $. Other
proofs are similar.
\begin{eqnarray}
\lambda^{P}_{\mathbf{c}_0} &=& \max_l
\left\{|P_{\mathbf{c}_0}(l)|, 1 \leq
l \leq m-1 \right\} \nonumber \\
&= & \max_l \left\{ |P_{\mathbf{s}_0}(l) + P_{\mathbf{s}_1}(l)|, 1
\leq l \leq \frac{m}{2}-1; |P_{\mathbf{s}_0,\mathbf{s}_1}(l) +
P_{\mathbf{s}_0,\mathbf{s}_1}(-l-1)|, 0 \leq l \leq \frac{m}{2}-1\right\} \nonumber \\
& \leq & \max_l \left\{ |P_{\mathbf{s}_0}(l)| +
|P_{\mathbf{s}_1}(l)|,
1 \leq l \leq \frac{m}{2}-1; 2|P_{\mathbf{s}_0,\mathbf{s}_1}(l)| , 0 \leq l \leq \frac{m}{2}-1\right\} \nonumber \\
&=& \max
\left\{\lambda^{P}_{\mathbf{s}_0}+\lambda^{P}_{\mathbf{s}_1},
2\lambda^{P}_{\mathbf{s}_0,\mathbf{s}_1}\right\} \nonumber
\end{eqnarray}
and
\begin{eqnarray}
S^{A}_{\mathbf{c}_0} &=& \sum_{l=1}^{m-1} |A_{\mathbf{c}_0}(l)|\nonumber \\
&=& \sum_{l=0}^{\frac{m}{2}-1} |A_{\mathbf{s}_0,\mathbf{s}_1}(l) +
A_{\mathbf{s}_0,\mathbf{s}_1}(-l-1)| + \sum_{l=1}^{\frac{m}{2}-1}
|A_{\mathbf{s}_0}(l) +
A_{\mathbf{s}_1}(l)| \nonumber \\
& \leq & \sum_{l=0}^{\frac{m}{2}-1}
|A_{\mathbf{s}_0,\mathbf{s}_1}(l)| + \sum_{l=-\frac{m}{2}+1}^{-1}
|A_{\mathbf{s}_0,\mathbf{s}_1}(l)| + \sum_{l=1}^{\frac{m}{2}-1}
|A_{\mathbf{s}_0}(l)| + \sum_{l=1}^{\frac{m}{2}-1}
|A_{\mathbf{s}_1}(l)| \nonumber \\ &=& S^{A}_{\mathbf{s}_0} +
S^{A}_{\mathbf{s}_1} + S^{A}_{\mathbf{s}_0,\mathbf{s}_1} \nonumber
\end{eqnarray}


\end{appendix}

\bibliographystyle{ieeetr}


\end{document}